\RequirePackage[l2tabu,orthodox]{nag}
\documentclass[copyright,hidelinks]{eptcs}
\pdfoutput=1

\usepackage[T1]{fontenc}
\usepackage{lmodern}
\usepackage{fixltx2e}
\usepackage{microtype}
\usepackage{xspace}
\usepackage{enumitem}

\makeatletter
\newcommand\etc{etc\@ifnextchar.{}{.\@}\xspace}
\newcommand\ie{i.e.\@\xspace}  

\makeatother

\usepackage{graphicx}

\newcommand{\inlinegraphic}[2]{
  \dimendef\grafheight=255\dimendef\grafvshift=254
  \grafheight=#1
  \grafvshift=-0.5\grafheight
  \advance\grafvshift by 0.5ex
  \raisebox{\grafvshift}{\includegraphics[height=\grafheight]{images/#2}\xspace}
}

\newcommand{\ninlinegraphic}[2][1.0]{
  \dimendef\grafheight=255\dimendef\grafvshift=254
  \setbox0 = \hbox{\scalebox{#1}{\includegraphics{images/#2}}}
  \grafheight=\the\ht0
  \grafvshift=-0.5\grafheight
  \advance\grafvshift by 0.5ex
  \raisebox{\grafvshift}{\includegraphics[height=\grafheight]{images/#2}\xspace}
}

\usepackage{latexsym}
\usepackage{amssymb}
\usepackage{amsmath} 
\usepackage{stmaryrd}
\usepackage{mathtools}
\usepackage{stackrel}

\usepackage{amsthm}
\newtheorem{theorem}{Theorem}[section]
\newtheorem{proposition}[theorem]{Proposition}
\newtheorem{lemma}[theorem]{Lemma}
\newtheorem{corollary}[theorem]{Corollary}
\theoremstyle{definition}\newtheorem{example}[theorem]{Example}
\theoremstyle{definition}
\theoremstyle{definition}\newtheorem{definition}[theorem]{Definition}
\theoremstyle{definition}
\theoremstyle{definition}\newtheorem{remark}[theorem]{Remark}
\theoremstyle{definition}
\newtheorem{notation}[theorem]{Notation}

\newcommand{\isomorphism}{\cong}

\ifx\numericids\undefined
\newcommand{\id}[1]{\ensuremath{\mathrm{id}_{#1}}}
\else
\newcommand{\id}[1]{\ensuremath{1_{#1}}}
\fi

\newcommand{\catP}{\ensuremath{{\cal P}}\xspace}

\newcommand{\catSet}{
\ensuremath{\mathbf{Set}}\xspace}

\newcommand{\Set}{\catSet}

\newcommand{\Pfn}{
\ensuremath{\mathbf{Pfn}}\xspace}
\newcommand{\Inj}{\ensuremath{\mathbf{Inj}}\xspace}

\usepackage{tikz-cd}
\newcommand{\lpushout}{\arrow[ur,phantom,"\urcorner",very near start]}
\newcommand{\rpushout}{\arrow[ul,phantom,"\ulcorner",very near start]}

\usepackage{dsfont}

\DeclareMathOperator{\cList}{\mathsf{CList}}%
\renewcommand{\int}[1]{\ensuremath{\mathrm{int}(#1)}}

\newcommand{\INC}{\ensuremath{\mathsf{inc}}\xspace}

\newcommand{\IMG}{\ensuremath{\mathsf{im}}\xspace}
\newcommand{\catrots}{\ensuremath{\mathcal{R}}\xspace}
\newcommand{\catgraf}{\ensuremath{\mathcal{G}}\xspace}
\newcommand{\catpregraf}{%
        \ensuremath{[\bullet\rightrightarrows\bullet,\Pfn]_\leq}\xspace}
\newcommand{\catbijgraf}{\ensuremath{\mathcal{B}}\xspace}

\usepackage{subcaption}
\usepackage{csvsimple}
\usepackage{multirow}
\usepackage{makecell}

\usepackage{hyperref}

\usepackage{tikzit}  

\newcommand{\inlinetikzfig}[2][1.0]{
  \dimendef\grafheight=255\dimendef\grafvshift=254
  \setbox0 = \hbox{\scalebox{#1}{\input{#2.tikz}}}
  \grafheight=\the\ht0
  \grafvshift=-0.5\grafheight
  \advance\grafvshift by 0.5ex
  \raisebox{\grafvshift}{\input{#2.tikz}}
}

\pgfdeclarelayer{background} 
\pgfdeclarelayer{foreground}
\pgfdeclarelayer{edgelayer}
\pgfdeclarelayer{nodelayer}
\pgfsetlayers{background,edgelayer,nodelayer,main,foreground} 

\tikzstyle{halfsize}=[x=0.5cm, y=0.5cm]
\tikzstyle{normalsize}=[]
\tikzstyle{doublesize}=[]

\tikzstyle{(null)}=[]
\tikzstyle{plain}=[]

\usepackage{tikzcircuits}
\usetikzlibrary{calc}
\usetikzlibrary{cd}

\tikzset{every picture/.style={line width=0.75pt}} 
\tikzset{directed edge/.style={postaction={decorate,decoration={markings,mark=at position 0.5 with {\arrow{>}}}}}}

\title{A Category of Surface-Embedded Graphs}

\author{Malin Altenm\"uller$^{1}$ \qquad \qquad \qquad Ross Duncan$^{1,2}$
\email{malin.altenmuller@strath.ac.uk \qquad ross.duncan@strath.ac.uk}
\institute{${}^1$Department of Computer and Information Sciences, University
  of Strathclyde,\\ 26 Richmond Street, Glasgow, G1 1XH, UK\\
  ${}^2$Quantinuum, Terrington House, 13--15 Hills Road, Cambridge
  CB2 1NL, UK}
}

\def\titlerunning{A Category of Surface-Embedded Graphs}
\def\authorrunning{M. Altenm\"uller \& R. Duncan}

\providecommand{\event}{ACT2022}

\begin{document}
\maketitle

\begin{abstract}
We introduce a categorical formalism for rewriting surface-embedded
graphs. Such graphs can represent string diagrams in a non-symmetric
setting where we guarantee that the wires do not intersect each
other. The main technical novelty is a new formulation of double
pushout rewriting on graphs which explicitly records the boundary of
the rewrite. Using this boundary structure we can augment these graphs
with a rotation system, allowing the surface topology to be incorporated.
\end{abstract}

\section{Introduction}
\label{sec:introduction}
String diagrams \cite{Selinger:2009aa} are a graphical formalism to
reason about monoidal categories. Equational reasoning in \emph{symmetric}
string diagrams can be implemented as graph or hyper-graph rewriting
subject to various side conditions to capture the precise flavour of
the monoidal category intended
\cite{Lucas-Dixon:2009yq,Lucas-Dixon-University-of-Edinburgh:2010fk,
  Dixon:2011fk,KZ:2015:aa,BGKSZ:esop17,BGKSZ:lics2016}.  We want to
use string diagrammatic reasoning for monoidal categories which are not
necessarily symmetric.  Informally, the lack of symmetry is often
stated as ``the wires cannot cross'' -- but what does that mean when the
string diagram is a graph or other combinatorial object?
Where is this ``crossing'' taking place?  To make sense of
this we must move beyond the situation where only the connectivity
matters and add some topological information.

In this paper we make two steps in that direction.  Firstly we borrow
a tool from topological graph theory -- rotation systems -- and use it
to define a category of graphs which are embedded in some surface.
Secondly, we introduce a new refinement of double pushout rewriting
\cite{Ehrig:2006ab} which is adapted to this category.  This
refinement was motivated by the desire to do rewriting on rotation
systems, however it works on conventional directed graphs equally
well, and removes many annoyances encountered when using standard
techniques from algebraic graph rewriting in string diagrams.  This is
an important step towards formalising non-symmetric string diagrams
and their rewriting theory.

Our motivation is also twofold.  From the abstract point of view,
non-symmetric string diagrams can capture a larger class of theories,
including both the symmetric case and the braided monoidal one.  A
more practical motivation comes from the area of quantum computing,
where string diagrams are often used to model quantum circuits
\cite{Coecke:2015aa}, their connectivity restrictions imposed by the
qubit architecture \cite{Alexander-Cowtan:2019aa} require a theory
without implicit SWAP gates, and can involve circuits defined on quite
complex surfaces.

Curiously, Joyal and Street's original work \cite{JS:1991:GeoTenCal1}
formalised monoidal categories as plane embedded diagrams, and used
the plane to carry the categorical structure.  Our work goes in the
opposite direction: to recover the topology from the combinatorial
structure.

\paragraph{Graph Embeddings}
Graphs can be drawn on surfaces, and the same graph be drawn
different ways on the same surface, as shown below.
\begin{center}
\includegraphics[width=0.3\textwidth]{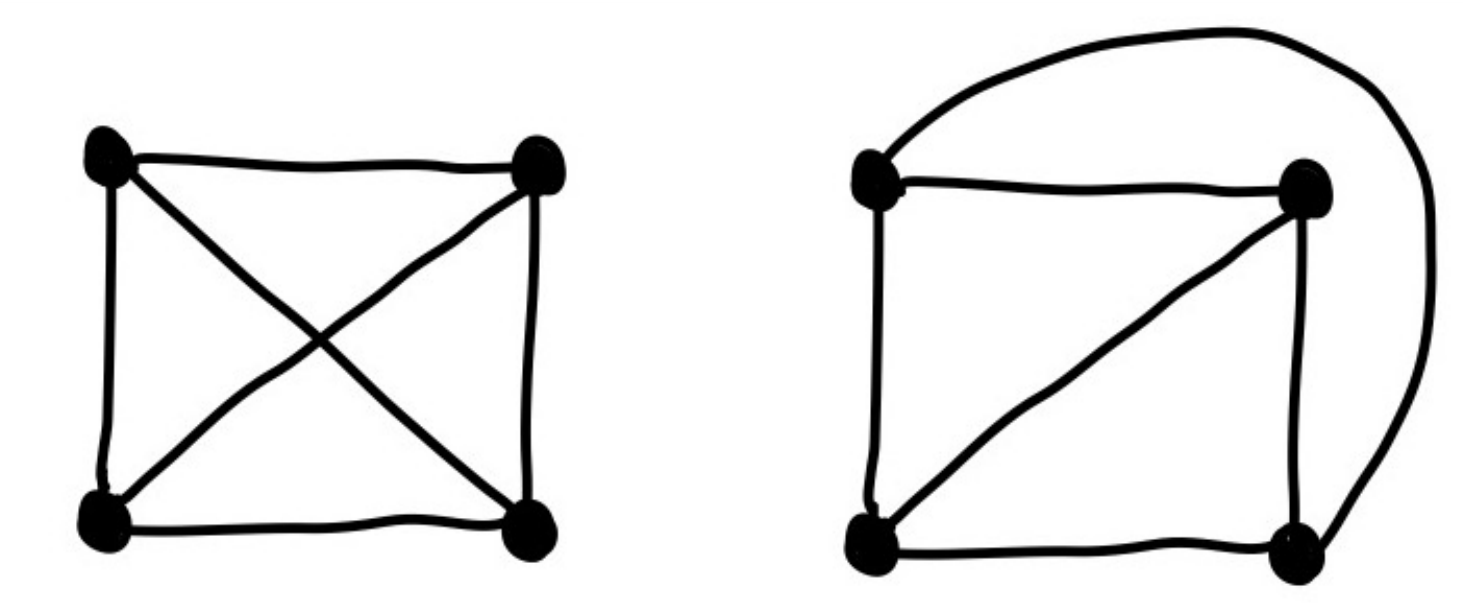}
\end{center}
If, like the one on the right, the drawing does not intersect itself
then it defines an \emph{embedding} of the graph into the surface.  If
a graph can be embedded in the plane (or equivalently surface of the
sphere) it is called \emph{planar}.  However in this work we will be
concerned with graphs with a given embedding into some closed compact
surface, which need not be the plane.

Dealing with lines and points as submanifolds of some surface (up to
homeomorphism) is quite unwieldy, so we use a combinatorial
representation of graph embeddings called \emph{rotation systems}.  A
rotation system imposes an order on the edges incident at vertex
(called a \emph{rotation}).  The rotation information at each vertex
is enough to fix the embedding of the graph into
some surface, as it defines the faces of the embedding uniquely. This is
 a well studied topic in graph theory and we refer to the literature for more details~\cite{Gross2001Topological-Gra}.
\begin{theorem}\label{thm:rots-give-embs}
  A rotation system determines the embedding of a connected graph into
  a minimal surface up to homeomorphism
  \cite{Heffter1891Uber-das-Proble,Edmonds1960A-combinatorial,Gross2001Topological-Gra}.
\end{theorem}

\noindent
Note that different rotation systems for the same graph may have
different minimal surfaces, which need not be the plane.

\paragraph{Boundary Graphs and Partitioning Spans}

When using string diagrams, graphs as usually defined are not the most
natural object; rather, we often think about \emph{open} graphs which
have ``half-edges'' or ``dangling wires'' which represent the domain
and codomain of the morphism in question.  The half-edges therefore
provide the interface along which morphisms compose, and also where
substitutions can be made in rewriting.  Unfortunately half-edges
don't work particularly well with double pushout rewriting,
necessitating various workarounds encoding the ``wires'' as special
vertices in a graph \cite{Dixon:2011fk} or hypergraph
\cite{BGKSZ:lics2016}.  This in turn leads to its own complications
when we consider the identity morphism, and other natural
transformations which are naturally ``all string''; equations which
should be trivial are no longer so.  Surface embedded graphs suggest a
different approach to this question.

Naively, when picturing a rewrite on a surface embedded graph, we
picture a disc-like region of the surface which is removed and
replaced.  The edges which cross the boundary of this region define
the interface and we naturally require that the removed disc and its
replacement should have the same interface.  From the outside, this
disc is homeomorphic to a point, so it can be treated as if it was a
vertex equipped with a rotation system.  However, the perspective from
inside and outside the region are completely equivalent, so we can
dually view the rest of the graph as a single vertex connected to the
interior of the disc.  We think of and draw a graph with boundary in
three different ways:
\begin{center}
\includegraphics[width=0.7\textwidth]{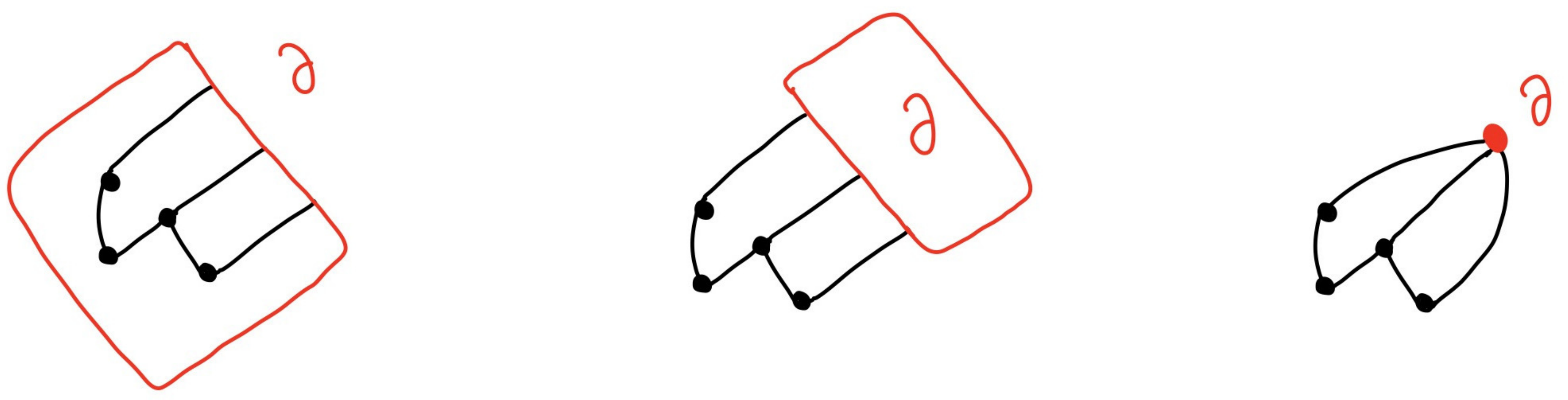}
\end{center}
On the left, the graph is depicted as a region of the surface with its outside
being the rest of the surface. In the middle, graph and its surrounding are
both regions of the surface, and on the right we have drawn the boundary as a
vertex with the interconnecting edges attached.

\begin{figure}[tb]
  \centering
  \includegraphics[width=0.7\textwidth]{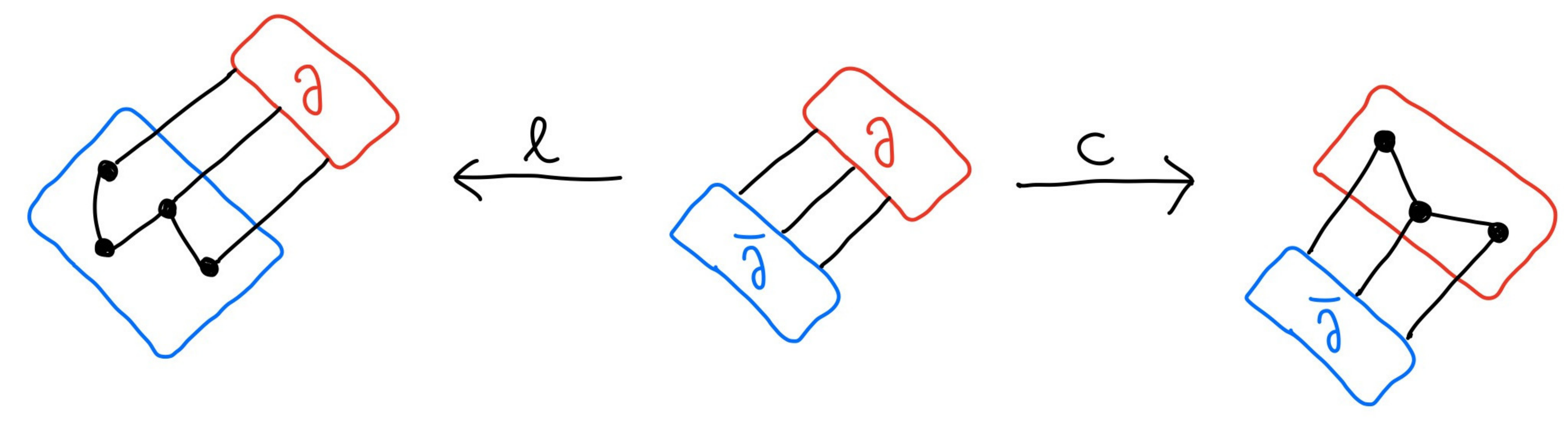}\\
  \includegraphics[width=0.7\textwidth]{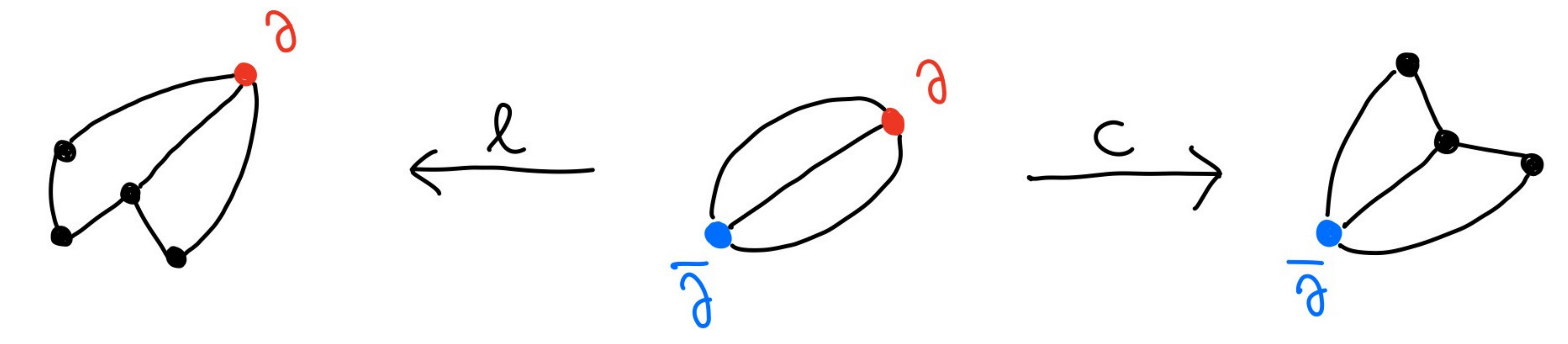}\\
  \caption{Example of a partitioning span, drawn in a region- and a vertex-style
    (edge directions are omitted for readability)}\label{fig:partion-span}
\end{figure}

This leads naturally to our notion of \emph{boundary graph}: we
contract both subgraphs on either side of the boundary to points,
leaving a two-vertex graph whose edges specify the connections
crossing the boundary.  Boundary graphs form the vertex of
\emph{partitioning spans}, which specify the whole graph as the two
parts, as shown in Figure~\ref{fig:partion-span}; the pushout of a
partitioning span is the original graph.

This formalism allow us to use a simple definition of graph, although
our morphisms are now built from partial functions, which introduces
some complications around the required injectivity properties to
preserve the type of the vertices, which is essential if these graphs
are to be interpreted as string diagrams.

\paragraph{Limitations}
The astute reader will have noted that
Theorem~\ref{thm:rots-give-embs} applies only to connected graphs.  To
specify an embedding of a disconnected graph a rotation system does
not suffice.  We would also need to take into account the relationship
between components and faces of the graph.  We have made no attempt to
do so here.

\section{A Suitable Category of Graphs}
\label{sec:suit-categ-graphs}

In this section we will introduce a category of directed graphs
without reference to any topological structure. The main
difficulty here is arriving at the correct notion of graph morphism:
our intent here is that the graphs represent terms in some monoidal
category -- \ie string diagrams -- and the morphisms represent
\emph{embeddings} of subterms.  This implies that certain structures
should be preserved which conventional graph rewriting does not worry
about.  Our choices here are also influenced by the variant of
double-pushout rewriting we will define in the next section.  In later
sections we will show how to incorporate the plane topology by adding
rotation systems.

A total graph is a functor
$G : (\bullet\rightrightarrows\bullet) \ \to \ \Set$. Concretely, such
a graph is a pair of sets $V$ and $E$, of \emph{vertices} and
\emph{edges} respectively, and a pair of functions $s$ and $t$
assigning \emph{source} and \emph{target} vertices to each edge.
\[
\begin{tikzcd}
  E \ar[r, shift left, "s"]
    \ar[r, shift right, "t"'] & V.
\end{tikzcd}
\]
In the functor category $[\bullet\rightrightarrows\bullet,\Set]$, a
morphism of graphs is a pair of functions
$f_V: V \rightarrow V',f_E : E \rightarrow E'$, such that the
following squares commute:

\begin{equation}\label{eq:edge-strict}
\begin{tikzcd}
  E \arrow[d, "s"'] \arrow[r, "f_E"]
& E' \arrow[d, "s'"] \\
  V \arrow[r, "f_V"']
& V'
\end{tikzcd}
\qquad\qquad
\begin{tikzcd}
  E \arrow[d, "t"'] \arrow[r, "f_E"]
& E' \arrow[d, "t'"] \\
  V \arrow[r, "f_V"']
& V'
\end{tikzcd}
\end{equation}
Sadly for us, this simple and elegant definition will not suffice.

We want to consider graph morphisms which can replace vertices with subgraphs,
and therefore forget these vertices, as shown below:
\begin{center}
\includegraphics[width=0.3\textwidth]{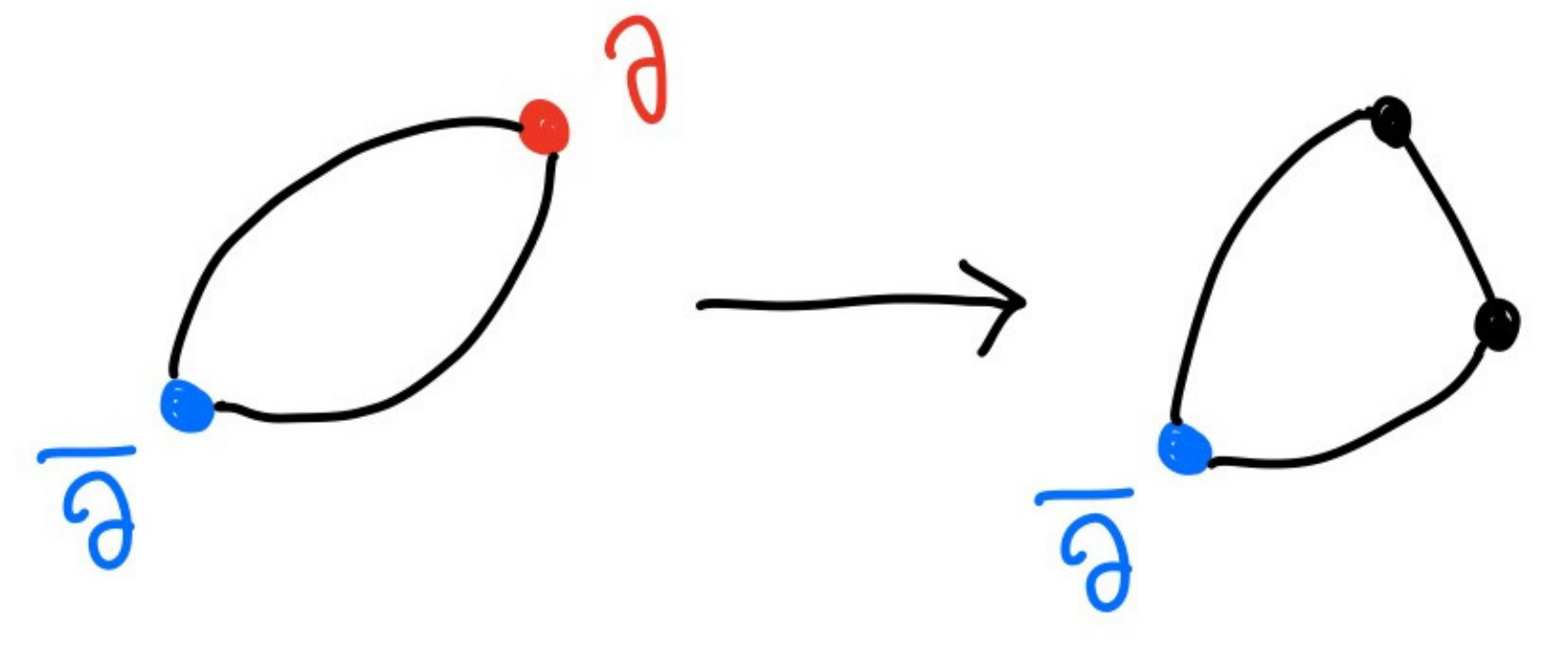}
\end{center}
To achieve this we could operate in a subcategory of
$[\bullet\rightrightarrows\bullet,\Pfn]$, the category of partial graphs and
maps, with only the total graphs as objects. However this is not quite
enough. Commutation of the naturality squares (\ref{eq:edge-strict}) in this
category is strict, meaning it includes equality of the domains of definition.
Therefore if a morphism forgets a vertex it must also forget all the incident
edges at that vertex. This is no use. We address this issue by using
the poset enrichment of \Pfn, and work in the category
$[\bullet\rightrightarrows\bullet,\Pfn]_\leq$ of functors and
\emph{lax} natural transformations:
\begin{equation}\label{eq:edge-lax}
\begin{tikzcd}
  E \arrow[d, "s"'] \arrow[r, "f_E"]
    \arrow[dr,phantom,"\leq"]
& E' \arrow[d, "s'"]
\\
  V \arrow[r, "f_V"', harpoon]
& V'
\end{tikzcd}
\qquad\qquad
\begin{tikzcd}
  E \arrow[d, "t"'] \arrow[r, "f_E"]
    \arrow[dr,phantom,"\leq"]
& E' \arrow[d, "t'"]
\\
  V \arrow[r, "f_V"', harpoon]
& V'
\end{tikzcd}
\end{equation}
The lax commutation allows the vertex component of a morphism to be
undefined at some vertex $v$ while its incident edges may be
preserved.  However, if an edge is ``forgotten'' then its source and
target vertices must also be so.  We'll need more, but let's take
$[\bullet\rightrightarrows\bullet,\Pfn]_\leq$ as our ambient category
for now.
\begin{proposition}\label{prop:pfn-pushouts}
The category \Pfn of sets and partial functions has pushouts.
\begin{proof}
Given a span
\begin{tikzcd}[cramped,sep=small]
    L & B  \arrow[l, "l"'] \arrow[r, "c"] & C
  \end{tikzcd}
, the elements of the pushout are the same as in for \Set, but restricted
  to a subset $B' \subseteq B$, with both $l(b)$ and $c(b)$
  defined for $b \in B'$. This is the only way the square commutes for elements
  in $B'$, and the universal property of the pushout can be derived from \Set.
\end{proof}
\end{proposition}
\begin{proposition}\label{prop:pinj-no-pushouts}
The category \Inj of sets and injective functions does not have
pushouts.
\begin{proof}
If pushouts in \Inj exist, they have to coincide with those of \Set. Consider
the span
\begin{tikzcd}[cramped, sep=small]
\{\ast\} & \emptyset \arrow[l] \arrow[r] & \{\ast\}
\end{tikzcd}, and commuting squares:
\begin{center}
\begin{tikzcd}
         & \{\ast\} \arrow[d] \arrow[ldd, "id" description, bend right] & \emptyset \arrow[l] \arrow[d]                               \\
         & {\{\ast,\ast\}} \arrow[ld, "m" description, dashed]          & \{\ast\} \arrow[l] \arrow[lld, "id" description, bend left] \\
\{\ast\} &                                                              &
\end{tikzcd}
\end{center}
In the square all morphisms are injective, but the mediating map out of the
pushout $m : \{\ast,\ast\} \rightarrow \{\ast\}$ is not.
\end{proof}
\end{proposition}
We would like to be able to accommodate two further properties in our notion of
graph morphism: Firstly, since vertices represent morphisms of a monoidal
category, their type should be preserved. Secondly, we want to specify when a
morphism is a graph \emph{embedding}, which requires an injectivity
property. Merely asking for injectivity of the vertex and edge component is not
enough though, our setup requires the edge component to be non-injective, \ie to
represent the identity morphism (or similar circumstances):
\begin{example}\label{ex:make-a-self-loop-example}
A graph morphism with a non-injective edge component:
\begin{center}
\includegraphics[width=0.3\textwidth]{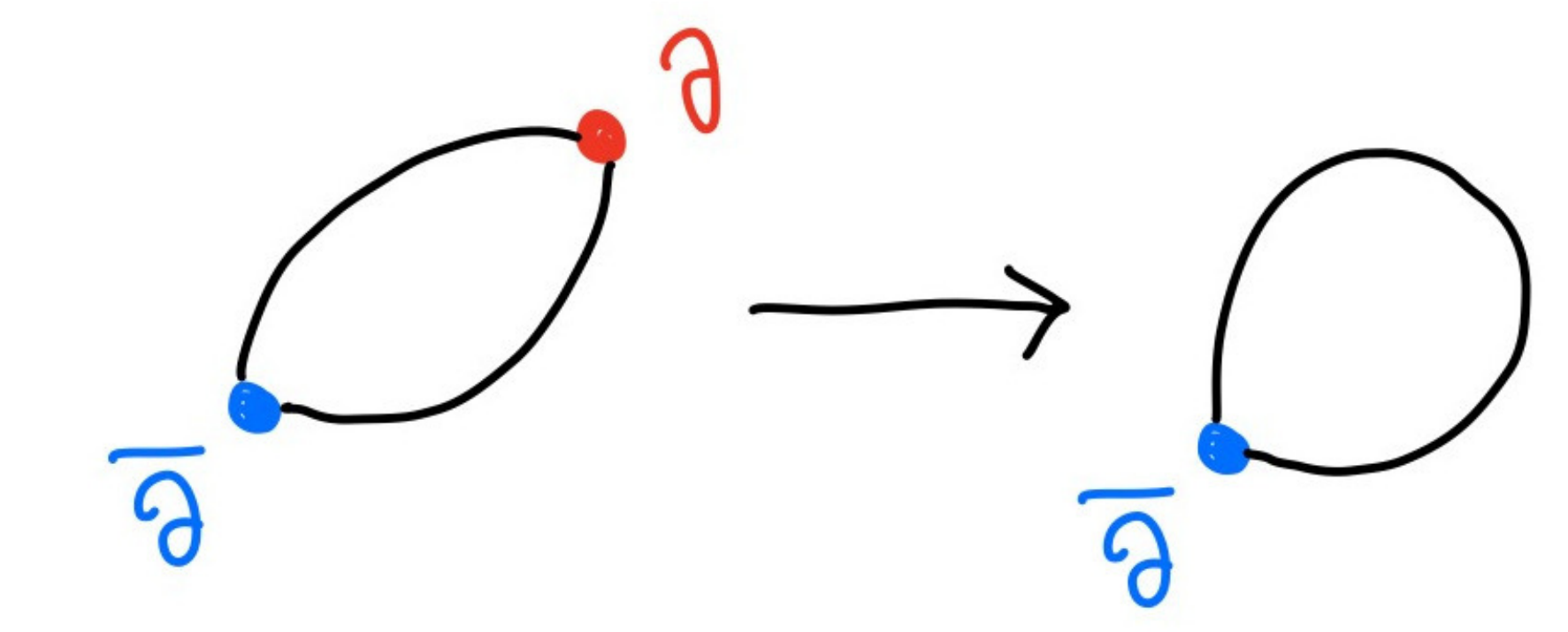}
\end{center}
\end{example}

Both of the above requirements turn out to be properties of the connection
points between vertices and their incident edges, called \emph{flags}:
\begin{definition}
  Given a graph $(V,E,s,t)$ its set of \emph{flags} is defined
  \[
    F = \{ (e,s(e)) : e \in E\} + \{ (e,t(e)) : e \in E\}
  \]
  Given a graph morphism $f:G\to G'$ there is an induced \emph{flag
    map}, $f_F : F \to F'$,
  \[
    f_F = (f_E \times f_V) + (f_E \times f_V)
  \]
\end{definition}

Note that the flag map is in general a partial map: it is undefined on $(e,
v)$, whenever $f_V$ is undefined on $v$.  Whenever $f_F$ is
injective we say that $f$ is \emph{flag injective}.

Flag injectivity allows edges to be combined but prevents a morphism
from decreasing a vertex degree in the process.  However, nothing said
so far forbids a morphism from increasing the degree of a vertex: we
require a notion of \emph{flag surjectivity}.  Given $f:G\to G'$, it
doesn't suffice to require the flag map $f_F$ to be surjective, since
in general $G'$ will contain more vertices than $G$, and hence more
flags.  The resulting definition is unfortunately unintuitive.

\begin{definition}\label{def:flag-surjective}
  Let $f:G\to G'$ be a morphism between two total graphs; we say that
  $f$ is \emph{flag surjective} if the two diagrams below commute
  laxly,
  \begin{equation}\label{eq:flag-surjectivity}
    \begin{tikzcd}
      V \arrow[r, "f_V", harpoon] \arrow[d, "s^{-1}"'] & V' \arrow[d,"s'^{-1}"]
      \\
      P(E) \arrow[ru, phantom, "\geq" description] \arrow[r, "P(f_E)"'] & P(E')
    \end{tikzcd}
    \qquad\qquad
    \begin{tikzcd}
      V \arrow[r, "f_V", harpoon] \arrow[d, "t^{-1}"'] & V' \arrow[d,"t'^{-1}"]
      \\
      P(E) \arrow[ru, phantom, "\geq" description] \arrow[r, "P(f_E)"'] & P(E')
    \end{tikzcd}
  \end{equation}
where $s^{-1}$ and $t^{-1}$ are the preimage maps of $s$ and $t$
respectively, and $P$ is the powerset functor.
\end{definition}
If a flag surjective morphism $f$ is defined on a vertex $v$, it will
  ensure that all edges attached to $v' = f_V(v)$ are in the image of $f_E$, thus no
  additional edges can be attached to $v'$ in the process. An example of a
  morphism which is not flag surjective can be found in Figure~\ref{fig:surj} in
  Appendix~\ref{app:examples}.\\
We'll call a morphism which is both flag injective and flag surjective
a \emph{flag bijection}.  This is quite a strong property; it's almost
enough to make the vertex map injective, but not quite.

\begin{lemma}\label{lem:almost-vertex-injective}
  Let $f:G\to G'$ be a flag bijection, and suppose that $f_V(v_1) =
  f_V(v_2)$ and both are defined; then $\deg {v_1} = \deg
  {v_2} = 0$.
  \begin{proof}
    Let $v' = f_V(v_1) = f_V(v_2)$; since $f$ is flag injective, the
    set of flags at $v'$ must contain (the image of) the disjoint
    union of the flags at $v_1$ and $v_2$; hence $\deg{v'} \geq
    \deg{v_1} + \deg{v_2}$. Since (by (\ref{eq:edge-lax})) $f_E$ is
    defined on all the flags at $v_1$, flag surjectivity implies that
    $\deg{v_1} \geq \deg{v'}$, and similarly for $v_2$.  Hence
    $\deg{v'} = \deg{v_1} = \deg{v_2} = 0$.
  \end{proof}
\end{lemma}

\begin{lemma}\label{lem:degree-preservation}
  Let $G$ and $G'$ be total graphs, and let $f:G\to G'$ be a flag
  bijection.  For all $v\in V$, if $f_V(v)$ is defined, then $f_E$
  defines a bijection between the flags at $v$ and those incident at
  $f_V(v)$; in consequence $\deg v = \deg{f_V(v)}$.
  \begin{proof}
    Let $v' = f_V(v)$.  The edges incident at $v$ are given by the
    disjoint union of $s^{-1}(v)$ and $t^{-1}(v)$, and likewise at
    $v'$.  Since $f$ is flag injective, $f_E$ is injective on the
    subset of flags defined by $v$.  Since $f$ is flag surjective all
    the flags at $v'$ are in the image of
    $f_E(s^{-1}(v))+f_E(t^{-1}(v))$.  Note that since $f_V(v)$ is
    defined then $f_E$ is defined for all $e \in s^{-1}(v)$ and all
    $e \in t^{-1}(v)$ by Eq.~(\ref{eq:edge-lax}).  Hence we have a
    bijection as required.
  \end{proof}
\end{lemma}

\begin{lemma}\label{lem:flag-bijections-compose}
Let $f:G\to H$ and $g:H\to J$ be flag bijections; then $g\circ f$ is a
flag bijection.
\begin{proof}
See Appendix \ref{app:proofs}.
\end{proof}
\end{lemma}

By the preceding lemma, and by observing that the identity is a
flag bijection, we may conclude that the flag bijections define a
wide subcategory of \catpregraf, which we will call \catbijgraf.

Example~\ref{ex:make-a-self-loop-example} suggests a confounding
special case: the vertex of a self loop can be forgotten. Here is another one:
\begin{example}\label{ex:self-loop-to-circle-eh}
  Let $G$ be the (unique) total graph with one vertex and one edge; let
  $G'$ be the (unique) partial graph with no vertices and a single edge.
  Define $f:G\to G'$ by $f_V= \emptyset$ and $f_E=\id{\mathbf 1}$.
  This is a valid flag bijection in \catbijgraf.
\begin{center}
\includegraphics[width=0.3\textwidth]{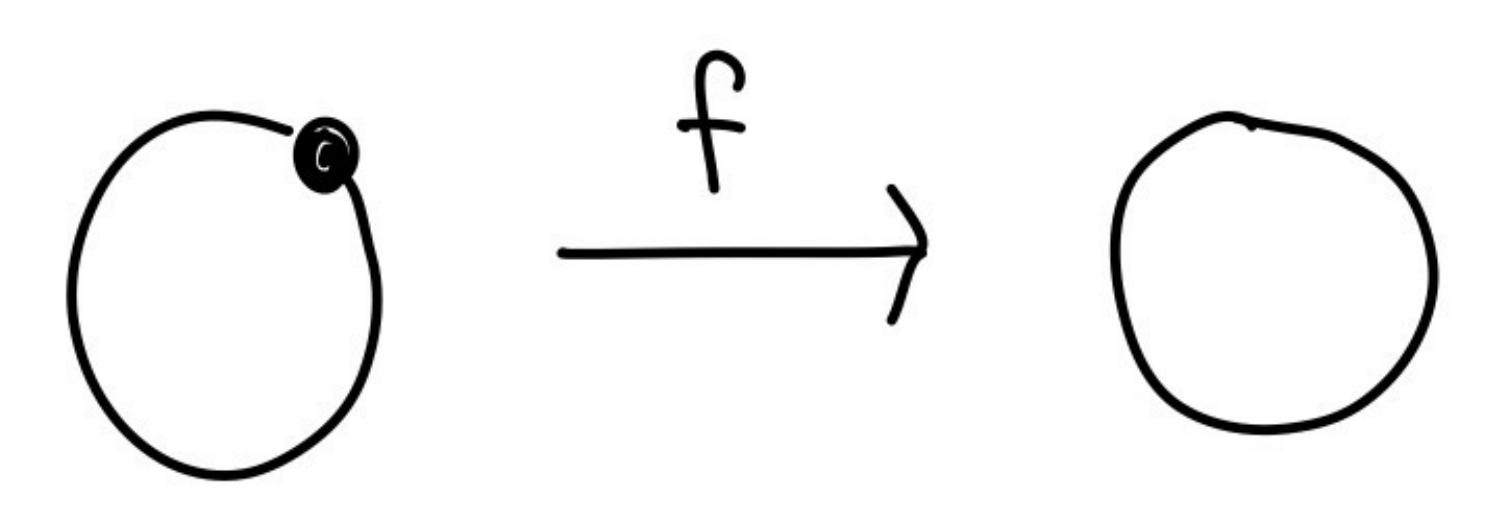}
\end{center}
\end{example}
While it is tempting to restrict to the subcategory defined by the
total graphs, and ban such monsters by fiat, they do occur quite
naturally in the rewrite theory we propose, albeit in quite restricted
circumstances. So they must be tamed.
To do so, we extend the definition of graph with
\emph{circles}: closed edges which have neither a source nor a target
vertex\footnote{This notion of graph has a long history; see, for
  example, the work of Kelly and Laplaza on compact closed
  categories \cite{KelLap:comcl:1980}.}.  Unfortunately the definition
of graph morphism will get more complex and the resulting category is
no longer a functor category, as we shall now see.

\begin{definition}\label{def:graph-with-circles}
  A \emph{graph with circles} is a 5-tuple $G=(V,E,O,s,t)$ where
  $(V,E,s,t)$ is a total graph and $O$ is a set of \emph{circles}.
  For notational convenience we define the set of \emph{arcs} as the
  disjoint union $A = E + O$.

  A morphism $f:G\to G'$ between two graphs with circles consists of
  two (partial) functions $f_V:V\rightharpoonup V'$ as above, and
  $f_A:A\to A'$, satisfying the conditions listed below.  Note that
  any such $f_A$ factors as four maps,
  \[
    \begin{array}{ccc}
      f_{E} : E \to E' & \ & f_{EO} : E \to O'\\
      f_{OE} : O \to E' & \ & f_{O} : O \to O'\\
    \end{array}
  \]
  The following conditions must be satisfied:
  \begin{enumerate}
  \item $f_A : A \to A'$ is total; \label{item:a-total}
  \item the component $f_{OE} : O \to E'$ is the empty
    function; \label{item:oe-empty}
  \item the pair $(f_V,f_E)$ forms a flag surjection between the
    underlying graphs in \catbijgraf.
    \label{item:flag-surjective}
  \end{enumerate}
If, additionally, the following three conditions are satisfied, we call the
morphism an \emph{embedding}:
\begin{enumerate}[resume]
   \item $f_V : V\rightharpoonup V'$ is injective; \label{item:v-inj}
   \item the component $f_{O}$ is injective; \label{item:o-injective}
   \item the pair $(f_V, f_E)$ forms a flag bijection between the underlying
     graphs. \label{item:flag-bijective}
\end{enumerate}
\end{definition}

\noindent
It's worth noticing that if some $f_A$ maps an edge $e$ to a circle, then
$f_E(e)$ is undefined, but $f_{EO}(e)$ is defined.  This, by the lax naturality
property, implies that $f_V$ is undefined on both $s(e)$ and $t(e)$. Various
examples and non-examples of morphisms and embeddings of graphs with circles can
be found in Appendix~\ref{app:examples}.

\begin{lemma}\label{lem:morphisms-compose}
  Defining composition point-wise, the composite of two morphisms of
  graphs with circles is again such a morphism. Additionally, if both morphisms
  are embeddings, their composition is an embedding as well.
  \begin{proof}
    See Appendix~\ref{app:proofs}.
  \end{proof}
\end{lemma}

\noindent
We finally have introduced all the necessary structure to define our
suitable category of graphs.

\begin{definition}
  Let \catgraf be the category whose objects are graphs with circles,
  and whose arrows are morphisms as per Definition \ref{def:graph-with-circles}.
\end{definition}

There is an obvious and close relationship between \catgraf and the
category of partial graphs and flag bijections, \catbijgraf.  We can
make this precise.

\begin{definition}\label{def:forgetful-functor}
  We define a forgetful functor $U : \catgraf \to \catbijgraf$ by
  \[
    \begin{tikzcd}
      U : (V,E,O,s,t)
      \arrow[r,mapsto]
      \arrow[d,"{U:(f_V,f_A)}"']
      \arrow[dr,phantom,"\mapsto"]
      & (V,E,s,t) \arrow[d,"{(f_V,f_E)}"]
      \\
      U : (V',E',O',s',t')
      \arrow[r,mapsto]
      & (V',E',s',t')
    \end{tikzcd}
  \]
\end{definition}

\begin{example}\label{ex:loop-to-circle-bis}
  Returning to Example \ref{ex:self-loop-to-circle-eh}, we see how
  this degenerate case fits in to the framework.  We start with $G$,
  the unique total graph with a single vertex and a single edge (and
  no circles).  There a single valid way to erase the vertex in
  \catgraf.

  Firstly observe that $G' =
  (\emptyset,\{e\},\emptyset,\emptyset,\emptyset)$ as in the earlier
  example is not an object in \catgraf.  However $G'' =
  (\emptyset,\emptyset,\{e\},\emptyset,\emptyset)$ is a valid graph,
  and the map $f:G\to G''$ which is undefined on the vertex and sends
  the edge to the circle is a valid morphism, indeed the only one.

  Finally observe that the image of $UG''$ is the empty graph and $Uf$
  is the empty function.
\end{example}

The term ``graph with circles'' is unacceptably cumbersome, so henceforth
we will simply say ``graph'' and refer to \catgraf as the
category of graphs.  In practice the circles are rarely important,
although we will devote a disappointingly large amount of this paper
to them.

\section{DPO Rewriting in the Suitable Category}
\label{sec:dpo-rewr-suit}

Double pushout rewriting \cite{Ehrig:2006ab} is an approach to
formalising equational theories over graphs by rewriting.  Each equation is
formalised as a rewrite rule $L \Rightarrow R$, and the substitution
$G[R/L]$ is computed via a double pushout as shown below.
\[
\begin{tikzcd}
   L \arrow[d, "m"']
&  B \arrow[d] \arrow[r, "r"]  \arrow[l, "l"']
&  R \arrow[d]
\\ G \lpushout
&  C \arrow[l]  \arrow[r]
&  H \rpushout
\end{tikzcd}
\]
The upper span embeds a \emph{boundary graph} $B$ into both $L$ and
$R$; ensures that both graphs have the same connectivity, and hence
that $R$ can validly replace $L$.  The map $m : L \rightarrow G$ is
the \emph{match}, an embedding of $L$ into $G$, which shows where the
rewrite will occur.  The first pushout square is completed by $C$, the
\emph{context graph}; it is basically $G$ with $L$ removed.  In the
DPO approach, $C$ is computed as a \emph{pushout complement}.  Finally
the graph $H = G[R/L]$ is the graph resulting from performing
the rewrite $L \Rightarrow R$ in $G$; it is computed as a pushout.

In the algebraic graph literature the notion of adhesive category
\cite{Lack:2003aa,Lack:2005aa} is commonly used, as DPO rewriting
behaves well in such categories.  However, adhesivity is not suitable
for our purposes, since the monomorphisms of \catgraf don't play any
special role in our formalism.  We will instead consider a specific case
of maps in the DPO diagram only, and in that context show the
existence of pushouts and the existence and uniqueness of pushout
complements, which are similar properties to those of an adhesive
categories.  The key to this approach is to recognise that $B$ and $C$
are in some sense partial graphs, as to a lesser extent are $L$ and
$R$; our handling of this partiality is one of the main novelties of
this paper.

\begin{notation}
  Almost every map in this section is an embedding of a small object
  into a larger one.  Wherever unambiguous to do so, we will treat
  these embeddings as actual inclusions so, for example, we may write
  $m_E(e) = e$ despite the fact that the domain and codomain of the map
  are different graphs.
\end{notation}

In our approach the graphs $L$ and $R$ that make up a rewrite rule
have an additional distinguished vertex, the \emph{boundary vertex}
$\partial$, which represents the rest of the world, from the
perspective of $L$ (or $R$).  The incident edges at $\partial$
represent the interface between $L$ and the rest of the graph it
occurs in.  The context graph $C$ also has a distinguished vertex, the
dual boundary $\bar\partial$  which represents its interface.  In our
formalism, the graph $B$ in the middle exists only to say that these
interfaces must be compatible.

\begin{definition}\label{def:boundary-graph}
  A \emph{boundary graph} is a graph with exactly two vertices,
  $\partial$ and $\bar\partial$ (called respectively the
  \emph{boundary} and \emph{dual boundary} vertices), where $s(e) \neq
  t(e)$ for all its edges $e$, and there are no circles.
\end{definition}

\begin{definition}\label{def:partitioning-span}
  A \emph{partitioning span} is a span
  \begin{tikzcd}[cramped,sep=small]
    L & B  \arrow[l, "l"'] \arrow[r, "c"] & C
  \end{tikzcd}
  in \catgraf, where $B$ is a boundary graph, the vertex component
  $l_V$ is defined on $\partial$ and undefined on $\bar\partial$ and, dually,
  $c_V$ is undefined on $\partial$ and defined on $\bar\partial$. Further, we
  require $l$ and $c$ to be embeddings.
\end{definition}

\begin{figure}[htb]
\centering
\begin{tikzcd}
L \arrow[d, "m"'] \arrow[r,leftarrow, "l"]
& B \arrow[d, "c"]
\\
G \arrow[r,leftarrow, "g"'] \lpushout
&
C
\end{tikzcd}
\qquad
\begin{tikzcd}
 \\
\includegraphics[width=0.37\textwidth]{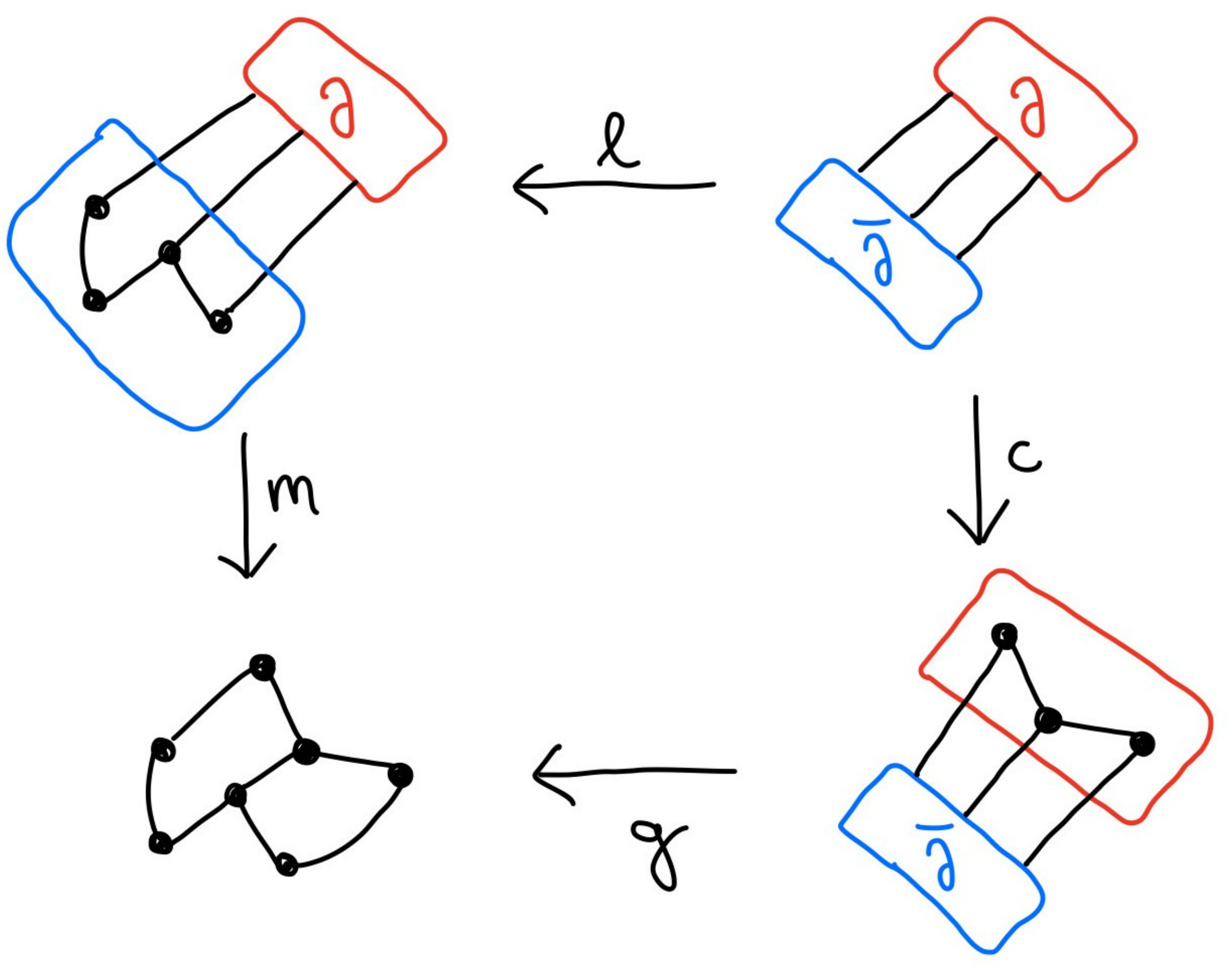}\\
\end{tikzcd}
\begin{tikzcd}
 \\
\includegraphics[width=0.37\textwidth]{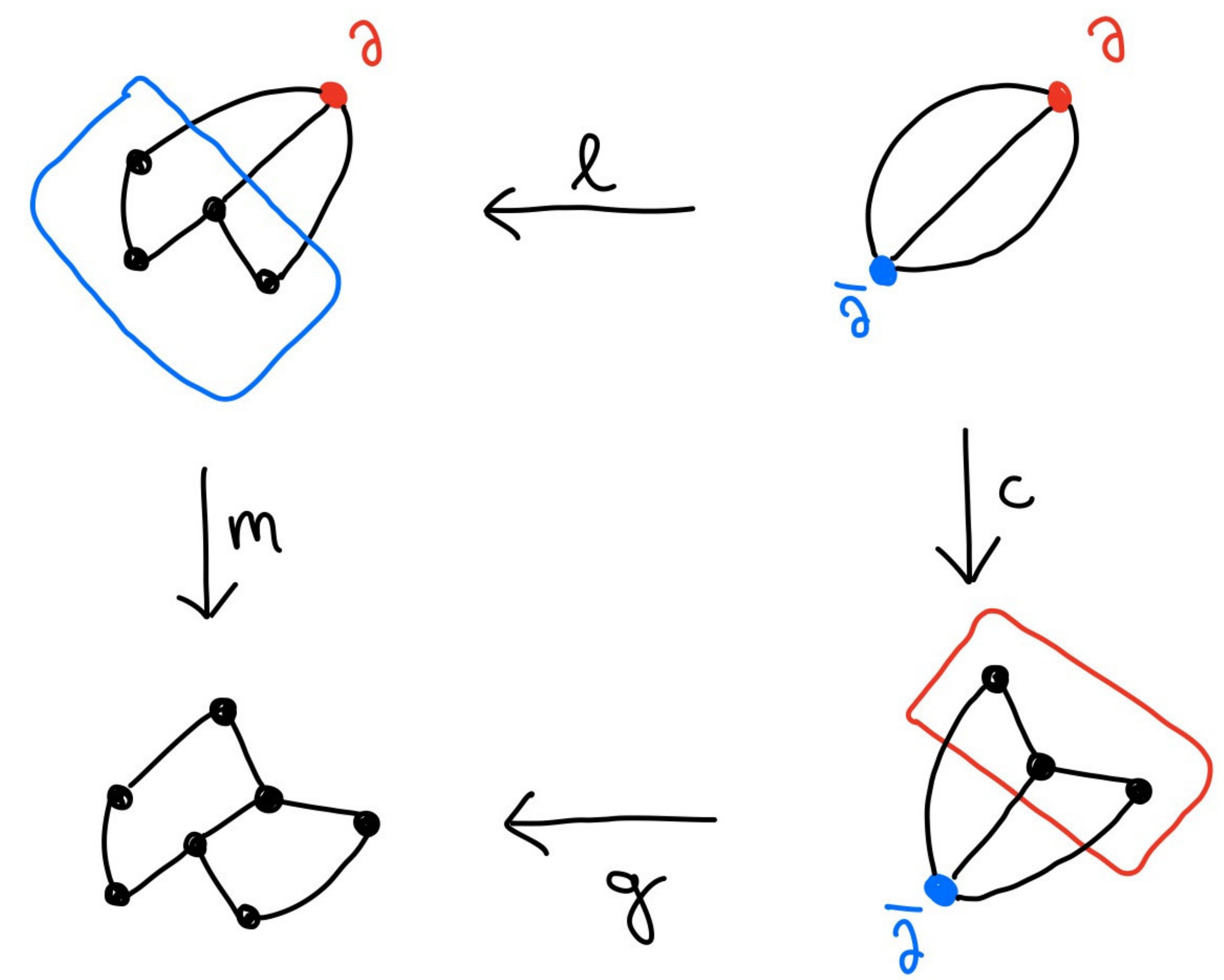}\\
\end{tikzcd}
\caption{Pushout of the partitioning span from Figure~\ref{fig:partion-span},
  drawn in two different ways}
\label{fig:ex-pushout}
\end{figure}
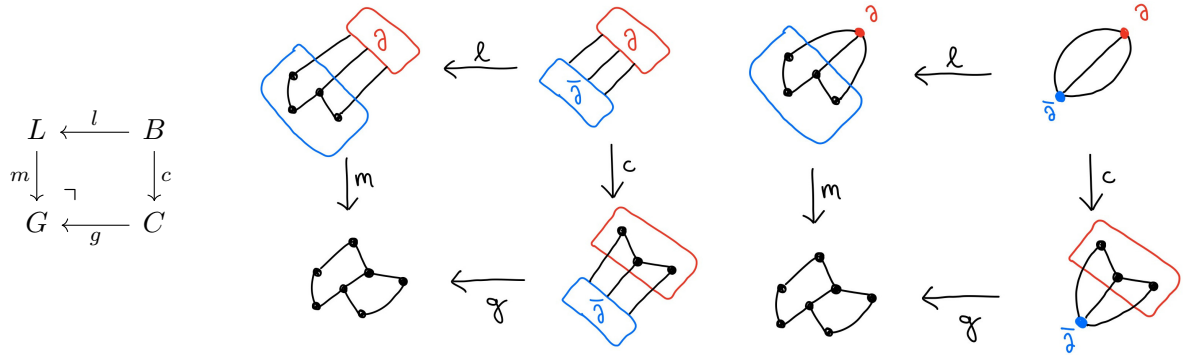

An example of a partitioning span and its pushout in \catgraf is
depicted in Figure~\ref{fig:ex-pushout}. The name \emph{partitioning
  span} arises from the fact that each of the maps out of the boundary
graph replaces one half of it. Hence each graph has two regions,
connected via the edges present in the boundary graph.

\begin{lemma}\label{lem:self-loop-creation}
  Let \begin{tikzcd}[cramped,sep=small]
    L & B  \arrow[l, "l"'] \arrow[r, "c"] & C
  \end{tikzcd}
  be a partitioning span and suppose that $e = l_E(e_1) = l_E(e_2)$ in $L$
  for distinct $e_1$ and $e_2$ in $E_B$.  Then $e$ is a self-loop at
  $\partial$ in $L$ and for all other $e_3 \neq e_1 \neq e_2$ we have $e\neq
  l_E(e_3)$.  The same holds mutatis mutandis for $C$.
  \begin{proof}
    By flag bijectivity all the flags at $\partial$ must be preserved,
    including distinct flags for $l_E(e_1)$ and $l_E(e_2)$.  By
    hypothesis these two edges are identified so necessarily
    $s_B(e_1) = \partial$ and $s_B(e_2) = \bar\partial$ or vice
    versa.  Hence $e$ is a self loop.  Suppose further that $l_E(e_3)
    = e$; then $l$ is not flag bijective, which is a contradiction.
  \end{proof}
\end{lemma}

Self-loops in partitioning spans indicate that the boundary is
connected back to itself without an intervening vertex.  This is
responsible for the failure of injectivity on edges and gives rise to
degeneracy when constructing pushouts.  We can study them using a dual
perspective.

\begin{figure}[thb]
\centering
\includegraphics[width=0.7\textwidth]{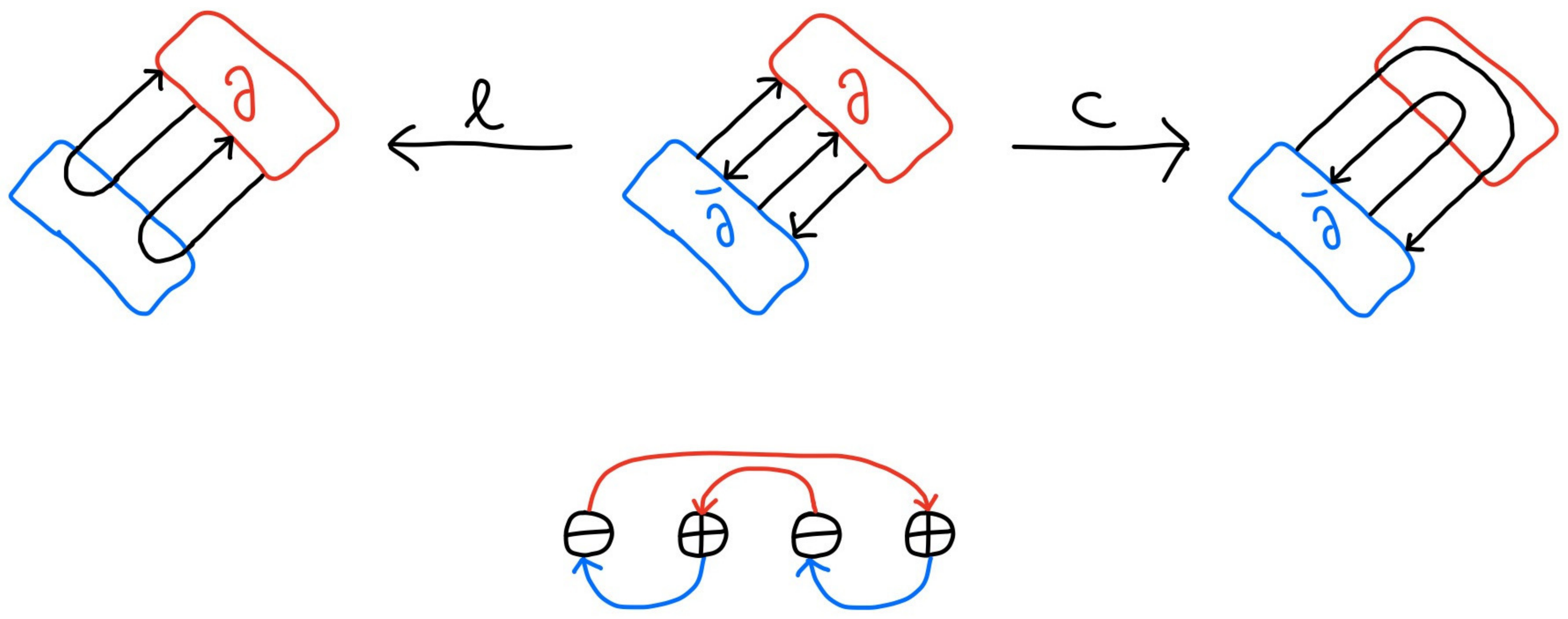}
\caption{Example of a partitioning span with its pairing graph}
\label{fig:pairing}
\end{figure}

\begin{definition}\label{def:pairing-graph}
  The \emph{pairing graph} for a partitioning span
    \begin{tikzcd}[cramped,sep=small]
    L & B  \arrow[l, "l"'] \arrow[r, "c"] & C
  \end{tikzcd}
  is a labelled directed graph whose vertices are $E_B$; each vertex receives
  a \emph{polarity}: $+$ if $s_B(e) = \partial$, $-$ if $s_B(e) = \bar\partial$.
  We draw a \emph{blue} edge between $e_1$ and $e_2$ if $l_E(e_1) = l_E(e_2)$
  i.e. if $e_1$ and $e_2$ form self-loop in $L$; similarly we draw a \emph{red}
  edge between $e_1$ and $e_2$ if they form a self-loop in $C$.  Blue edges are
  directed from positive to negative polarity; red edges the reverse.
\end{definition}
An example of a pairing graph is shown in Figure~\ref{fig:pairing}.
The pairing graph is always bipartite: it's immediate from the
definition that vertices of the same polarity are never connected.
Further, due to Lemma~\ref{lem:self-loop-creation}, each vertex can
have a maximum of one edge of each colour incident to it.  In
consequence every connected component is just a path, possibly of length
zero, possibly a cycle.  From these properties, we have the following
immediate corollary.

\begin{corollary}\label{cor:pgraph-comps-are-paths-in-B}
  Let \catP be the pairing graph of the partitioning span
  \begin{tikzcd}[cramped,sep=small]
    L & B  \arrow[l, "l"'] \arrow[r, "c"] & C
  \end{tikzcd}; then each connected
  component $p$ of \catP determines an edge-disjoint path on $B$.  For
  those components which are not cycles, if the first vertex of $p$ is
  positive, then the path starts at $\partial$; if negative the path
  starts at $\bar\partial$.  Conversely, if the last vertex of $p$ is
  positive, the path ends at $\bar\partial$ and vice versa.
\end{corollary}

\noindent
The reader may already suspect that when we form the pushout of a
partitioning span, the components of the pairing graph determine which
edges in $B$ will be identified.  This is indeed the case; it forms an
intermediate result (Lemma~\ref{lem:path-in-B}) in the proof of the next theorem.

\begin{theorem}\label{thm:pushouts}
In \catgraf, pushouts of partitioning spans exist. Further, the maps into the
pushout are embeddings.
\begin{proof}
  See Appendix~\ref{app:proofs}.
\end{proof}
\end{theorem}

Since pushouts of partitioning spans are the basis of the rewrite theory we
wish to pursue, for the rest of the paper the term ``pushout'' should
be understood to imply ``of partitioning span''.

We now move on to the other required ingredient for DPO rewriting:
pushout complements.  Just as we did with partitioning spans and
pushouts, we will introduce a specific kind of embedding for which the
complement must exist.

\begin{definition}\label{def:boundary-embedding}
A \emph{boundary embedding} is a pair of maps
\begin{tikzcd}[cramped,sep=small]
B \arrow[r,"l"] & L \arrow[r,"m"] & G
\end{tikzcd}
in \catgraf, where $B$ is a boundary graph, where :
(i) $l_V(\partial)$ is defined but $l_V(\bar\partial)$ is undefined;
and (ii) $(m_V\circ l_V)(\partial)$ is undefined. Further, $L$ has to be a
connected graph, and $m$ an embedding.
\begin{figure}[tb]
  \centering
  \includegraphics[width=0.7\textwidth]{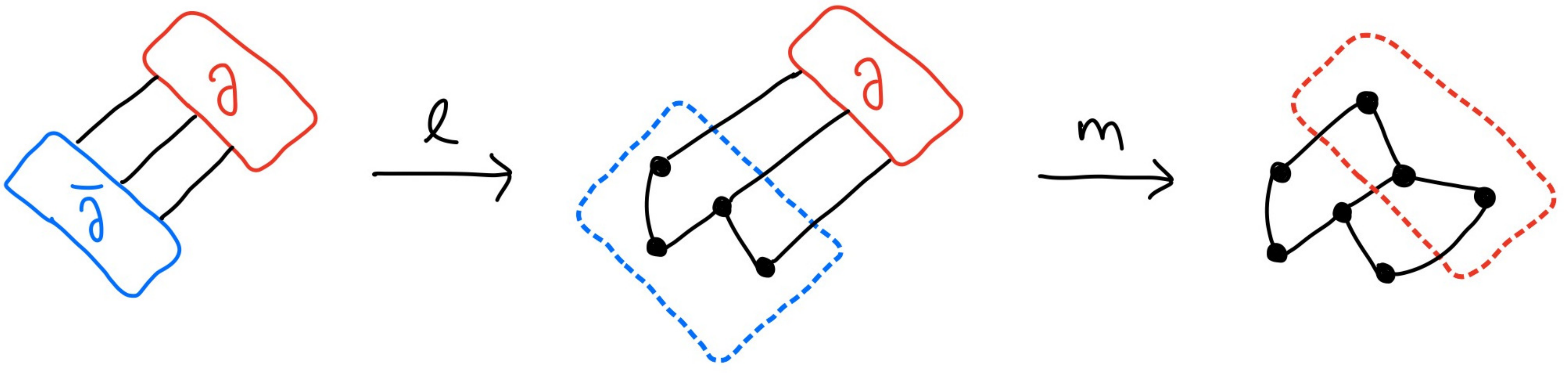}\\
  \caption{Example of a boundary embedding}
  \label{fig:boundary-emb}
\end{figure}
\end{definition}

\begin{figure}
\centering
\includegraphics[width=0.9\textwidth]{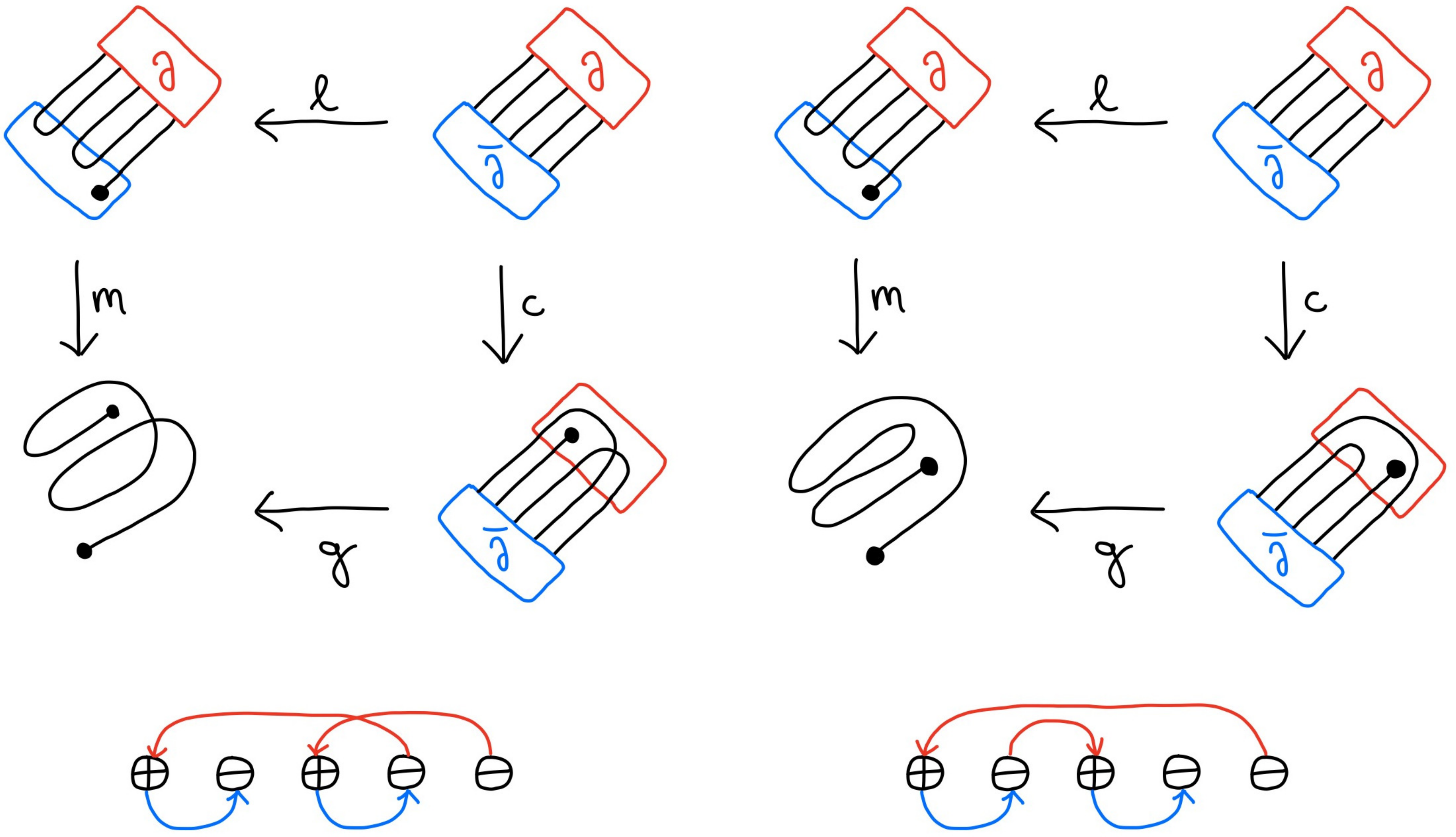}
\caption{Two different solutions to the same re-pairing problem, together with
  the corresponding pairing graphs.}
\label{fig:re-pairing}
\end{figure}

\begin{definition}\label{def:re-pairing}
  Given a boundary embedding
  \begin{tikzcd}[cramped,sep=small]
    B \arrow[r,"l"] & L \arrow[r,"m"] & G
  \end{tikzcd}
  we can immediately construct half a pairing graph \catP, consisting
  of only the blue edges using the mapping $l:B\to L$.  The
  \emph{re-pairing problem} is to construct the other half (the red
  edges) so that the connected components map to the edges of $G$
  (cf. Lemma~\ref{lem:path-in-B}).  See
  Figure~\ref{fig:re-pairing} for examples.
\end{definition}

\begin{lemma}\label{lem:repairing-has-a-solution}
Given a boundary embedding
\begin{tikzcd}[cramped,sep=small]
B \arrow[r,"l"] & L \arrow[r,"m"] & G
\end{tikzcd}
a solution to the re-pairing problem always exists, but it is not
necessarily unique.
\begin{proof}
  See Appendix~\ref{app:proofs}.
\end{proof}
\end{lemma}

\begin{theorem}\label{th:pushout-complements-ex}
  In \catgraf, pushout complements of boundary embeddings exist, and
  give rise to partitioning spans.
\begin{proof}
  We'll use the boundary embedding
  \begin{tikzcd}[cramped,sep=small]
    B \arrow[r,"l"] & L \arrow[r,"m"] & G
  \end{tikzcd}
  to  construct the complement $C$ such that
  \begin{tikzcd}[cramped,sep=small]
    L  & B \arrow[l,"l"'] \arrow[r,"c"] & C
  \end{tikzcd}
  is a partitioning span, and show that $G$ is indeed the pushout of
  this span.

  Let $C$ have vertex set $V_C = (V_G\setminus
  V_L)+\{\bar\partial\}$.  We'll construct the edge set, and the
  source and target maps, in three steps.
  \begin{enumerate}
  \item Let $E_C$ contain all the edges of the induced subgraph of $G$
    defined by the vertices $V_C$, and define the source and target
    maps on those edges correspondingly.
  \item Let $O_C$ contain $O_G \setminus m_O^{-1}(O_G)$.
  \item Finally we add the edges between $\bar\partial$ and the rest
    of the graph, and simultaneously define the map $c:B\to C$.  Let
    \catP be a solution to the re-pairing problem given by
    \begin{tikzcd}[cramped,sep=small]
      B \arrow[r,"l"] & L \arrow[r,"m"] & G
    \end{tikzcd}.
    If in \catP there is a red edge between $e_1$ and $e_2$ in create
    a self-loop $e$ at $\bar\partial$ and set $c(e_1) = c(e_2) = e$.
    If there is any vertex $e$ in \catP which has no incident red
    edge, add $e$ to $E_C$; if its polarity is positive set
    \[
      s_C(e) = (s_G \circ m_E \circ l_E)(e)
      \qquad\qquad
      t_C(e) = \bar\partial
    \]
    and if the polarity is negative, the source and target are
    reversed.  We define $c_E(e) = e$.
  \end{enumerate}
  The resulting span
  \begin{tikzcd}[cramped,sep=small]
    L  & B \arrow[l,"l"'] \arrow[r,"c"] & C
  \end{tikzcd}
  is evidently partitioning, and by construction has $G$ as its
  pushout, as a consequence of
  Lemma~\ref{lem:path-in-B}
\end{proof}
\end{theorem}

\begin{theorem}\label{th:complements-unique}
In \catgraf, pushout complements are unique up to the solution of the re-pairing
problem.
\begin{proof}
  Suppose that both
  \begin{tikzcd}[cramped,sep=small]
    B \arrow[r,"c"] & C \arrow[r,"g"] & G
  \end{tikzcd}
  and
  \begin{tikzcd}[cramped,sep=small]
    B \arrow[r,"c'"] & C' \arrow[r,"g'"] & G
  \end{tikzcd}
  are pushout complements for the boundary embedding
  \begin{tikzcd}[cramped,sep=small]
    B \arrow[r,"l"] & L \arrow[r,"m"] & G
  \end{tikzcd}.
  Observe that given the boundary embedding, a solution to the
  re-pairing problem determines the map $c:B\to C$ and vice versa.
  Let's assume for now a that $\IMG(c) = \IMG(c')$ and hence they both
  correspond to the same pairing graph.

  Since $m$ is an embedding, it follows that every part of $C$ not in
  $\IMG(c)$ is preserved isomorphically in $G$, and similarly for
  $C'$.  Since we have assumed $\IMG(c) = \IMG(c')$ this implies that $C
  \simeq C'$.

  Further, observe that different solution of the re-pairing also have
  the same number of edges, and hence produce the same number of self
  loops at $\bar\partial$.  Hence the difference between different
  solutions is just the labels on the edges incident at $\bar\partial$.
\end{proof}
\end{theorem}

\section{A Category of Rotation Systems}
\label{sec:categ-rotat-syst}

Despite some suggestive illustrations, up to this point we have
operated in a purely combinatorial setting, but now we introduce some
topological information in the form of rotation systems.  A rotation
system for a graph determines an embedding of the graph into a surface
by fixing a cyclic order of the incident edges, or more precisely the
flags, at every vertex.

We augment our category of graphs with this extra structure, in the
form of \emph{cyclic lists of flags} for each vertex, and strengthen
the property of flag surjectivity
(Equation~\ref{eq:flag-surjectivity}).  The requisite categorical
properties for DPO rewriting will follow more or less immediately from
those of the underlying category of directed graphs.

\begin{definition}\label{def:clist}
  Let $\cList : \catSet \to \catSet$ be the functor where $\cList X$
  is the set of circular lists whose elements are drawn from $X$.
\end{definition}

\begin{definition}\label{def:rotation-system-bis}
  A \emph{rotation system} $R$ for a graph with circles $(V,E,O,s,t)$ is
  a total function $\INC : V \to \cList E$ such that :
  \begin{itemize}
  \item $e \in \INC(s(e))$
  \item $e \in \INC(t(e))$
  \item $t^{-1}(v) + s^{-1}(v) \isomorphism \INC(v)$ (when considering $\INC(v)$
    as a set)
  \end{itemize}
We call $\INC(v)$ the \emph{rotation} at $v$.
\end{definition}

\noindent
Note that $\INC(v)$ is actually a cyclic ordering on the set of flags at $v$.

\begin{definition}\label{def:rotsys-morphism}
  A homomorphism of rotation systems $f:R\to R'$ is a \catgraf-morphism
  $(f_A,f_V)$ between the underlying graphs, satisfying the following additional
  condition.
  \begin{equation}\label{eq:rot-preserve}
    \begin{tikzcd}
      V \arrow[d, "\INC"'] \arrow[r, "f_V", harpoon]
      & V' \arrow[d, "\INC'"] \\
      \cList E \arrow[r, "\ \cList f_E\ "']
      \arrow[ru, phantom, "\geq" description]
      & \cList E'
    \end{tikzcd}
  \end{equation}
\end{definition}
This condition requires the preservation of the edges ordering on
vertices where $f_V$ is defined; it implies flag surjectivity
(Equation~\ref{eq:flag-surjectivity}).  Morphisms therefore
either preserve a vertex with its rotation exactly, or forget about
it.

\begin{definition}\label{def:rotsys-cat}
Let \catrots be the category whose objects are tuples $(V,E,O,s,t,\INC)$ where
$(V,E,O,s,t)$ is an object of the category of graphs, \catgraf (see
Def.~\ref{def:graph-with-circles}) and \INC is a rotation system for this
graph. The morphisms of \catrots are homomorphisms of rotation systems.
\end{definition}

\noindent
There is an evident forgetful functor $U':\catrots \to \catgraf$; this
is especially clean since the morphisms of \catrots are just
\catgraf-morphisms which satisfy an additional condition.  Further,
since we demand require the \INC structure to be preserved exactly,
pushouts and complements are very easily defined here.

\begin{definition}\label{def:rotsys-prop}
  In \catrots, objects $B$,  spans
  \begin{tikzcd}[cramped, sep=small]
    L & B  \arrow[l, "l"'] \arrow[r, "c"] & C
  \end{tikzcd}
  and composites
  \begin{tikzcd}[cramped,sep=small]
    B \arrow[r,"l"] & L \arrow[r,"m"] & G
  \end{tikzcd}
  are respectively \emph{boundary graphs}, \emph{partitioning spans},
  and \emph{boundary embeddings} if their underlying graphs in
  \catgraf satisfy those definitions (respectively Definitions
  \ref{def:boundary-graph}, \ref{def:partitioning-span}, and
  \ref{def:boundary-embedding}).
\end{definition}

\begin{lemma}\label{lem:rotsys-pushouts}
In \catrots pushouts of partitioning spans exist.
\begin{proof} The pushout candidate is the one in the underlying category (see
  Theorem~\ref{thm:pushouts}), together with the rotation system:
\[
\INC_G(v) = \left \{
          \begin{array}{ll}
          \INC_C(v), & \text{if } v \in V_C\\
          \INC_L(v), & \text{if } v \in V_L
          \end{array}
\right.
\]
The vertex set of the pushout is the disjoint union of vertices from both input
graphs, $V_G = (V_L + V_C)\setminus V_B$. Therefore, by the mediating map from
Theorem~\ref{thm:pushouts}, $\INC_G$ is indeed the pushout of the rotations.
\end{proof}
\end{lemma}

\begin{lemma}\label{lem:rotsys-complements-ex}
In \catrots pushout complements of boundary embeddings exist, and are
unique up to the solution of the re-pairing problem.
\begin{proof}
  This follows from the underlying construction in \catgraf; see
  Theorem \ref{th:complements-unique}.  Note that the rotation for
  every vertex of $C$ is determined by either those of $G$ or of $B$,
  so there is no choice about the additional structure.
\end{proof}
\end{lemma}


\begin{remark}\label{rem:po-complement-not-so-unique}
  We must sound a cautionary note about the ``up to'' in the preceding
  statement.  While in \catgraf pushout complements that arise from
  different pairing graphs are essentially the same, this is not so in
  \catrots.  Since the rotation around $\bar\partial$ is preserved
  exactly by $c:B\to C$, different choices for which edges to merge as
  self loops will result in different local topology at
  $\bar\partial$.  In particular it can happen that a re-pairing
  problem can have planar and non-planar solutions; see
  Figure~\ref{fig:re-pairing} for an example.
\end{remark}

With that caveat noted, since \catrots has pushouts and their
complements, specialised to the setting where the rewrite rules
explicitly encode the connectivity at their boundary, we can use it as
a setting for DPO rewriting of surface embedded graphs.

\begin{remark}
  As illustrated in Figure~\ref{fig:re-pairing}, we have adopted a
  particular convention for drawing the pairing graphs: the vertices
  are drawn in a row, with the red edges above and the blue edges
  below.  If the vertices are drawn in an order compatible with
  $\INC_B(\partial)$ then the blue edges (partly) reproduce the local
  topology at $\partial$ in $L$.  Any edge crossings imply the region
  around $\partial$ is not planar.  This is sufficient but not
  necessary for $L$ to be non-planar.  Isomorphic statements can be made
  for $\bar\partial$ in $C$.
\end{remark}

\section{On Planarity}
\label{sec:planarity}

Since the graphs of \catrots are equipped with rotation systems they
carry information about their topology along with them.  As the
previous section showed, \catrots admits DPO rewriting, but we might
ask for more, for example, to maintain a topological invariant.
Concretely, we might ask: if $L$, $R$, and $G$ are all plane
embedded is $G[R/L]$ also plane?  We have already seen, in
Remark~\ref{rem:po-complement-not-so-unique} above, that the re-pairing
problem can have topologically distinct solutions.

Focussing on plane graphs, it's possible that the re-pairing problem has
distinct plane solutions, for example :
\begin{center}
\includegraphics[width=0.7\textwidth]{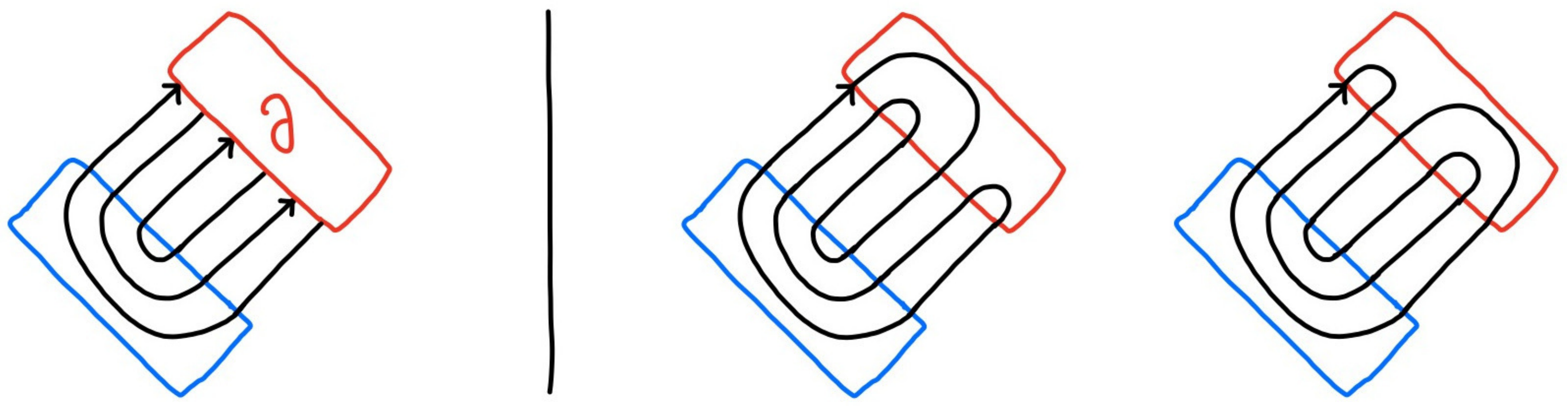}
\end{center}
It may also occur that there is no plane solution.  Consider a rewrite rule:\\
\begin{center}
  \includegraphics[width=0.25\textwidth]{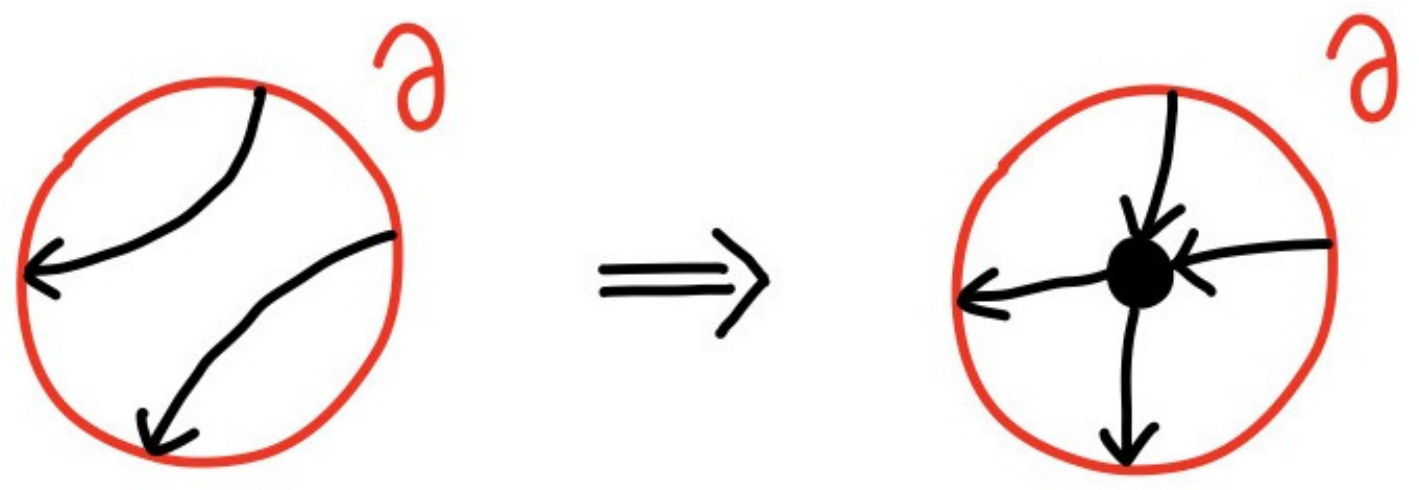}
\end{center}
This is a legitimate rewrite rule for plane graphs, and expanding it into a span
for the top of a double pushout diagram makes sense.
\begin{center}
  \includegraphics[width=0.6\textwidth]{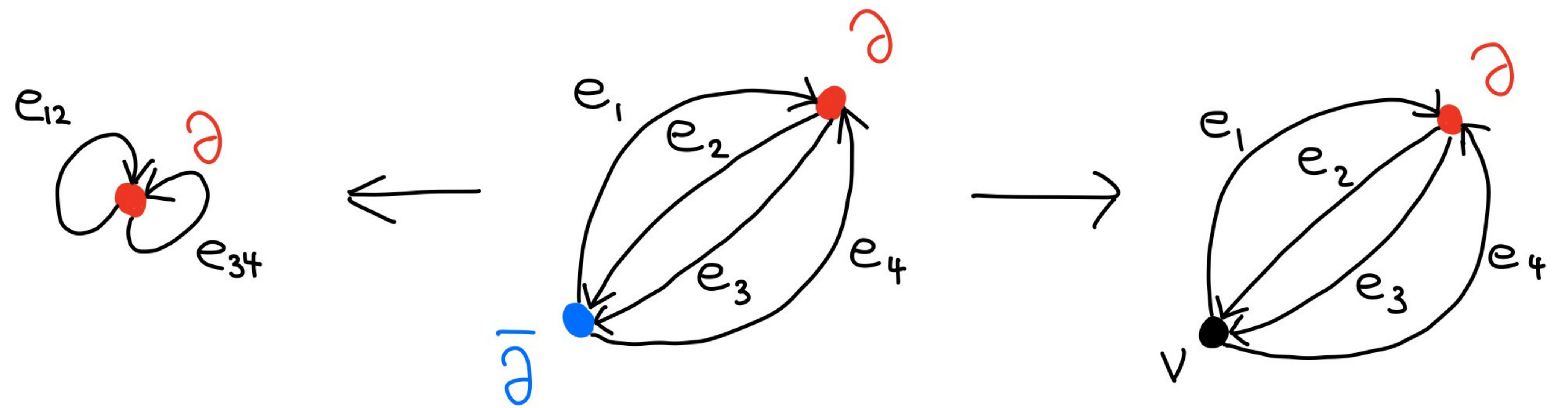}
\end{center}
Further it's clear that the left-hand side can be embedded into a
circle, which is trivially plane.  However when we apply the rewrite
something goes wrong.
 \begin{center}
 \includegraphics[width=0.6\textwidth]{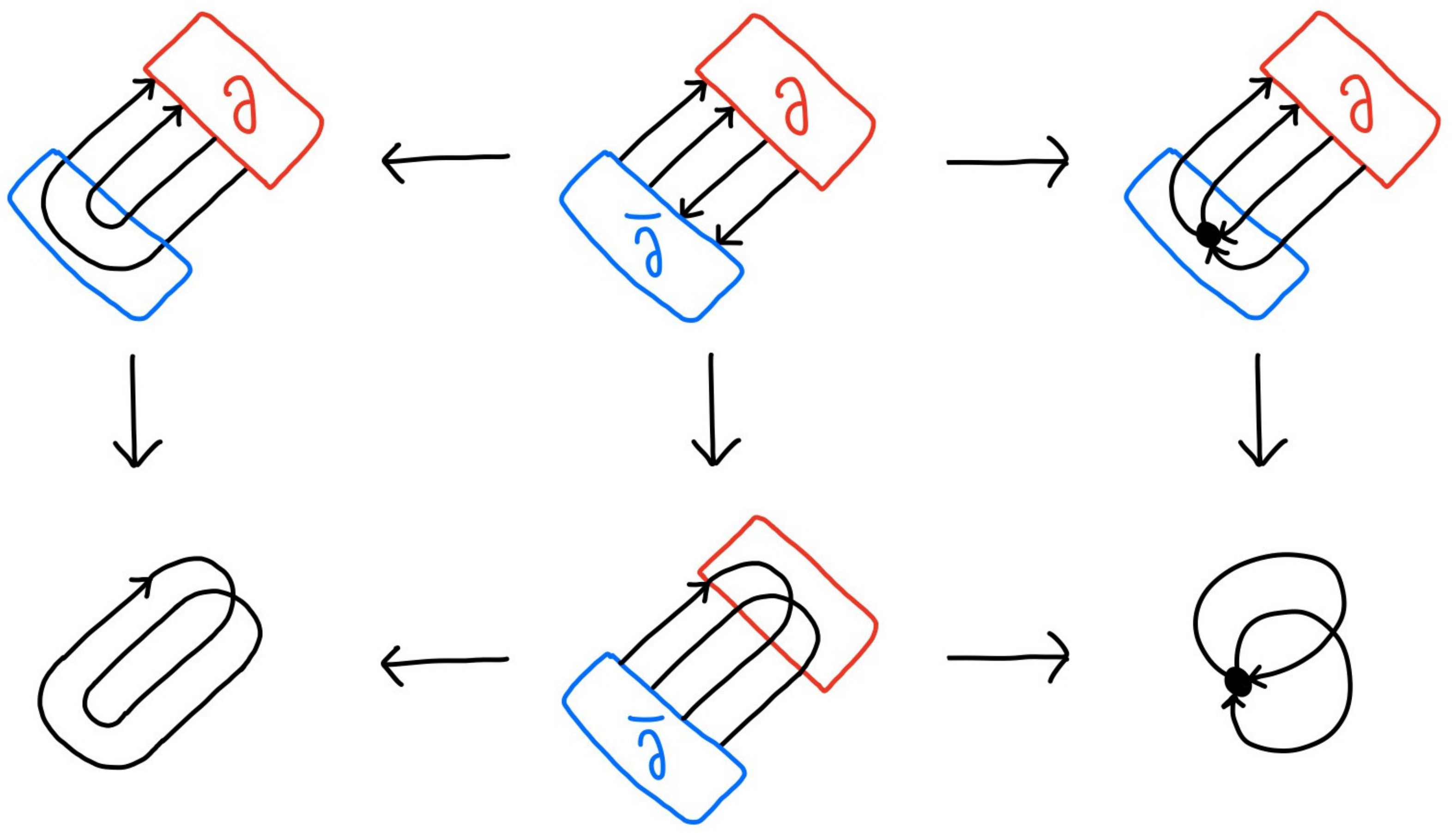}
 \end{center}
In this example we match the left hand side of the rule to the graph
with one circle and no vertices, and compute the complement and the
result of substituting the right hand side into the context graph.
When computing the pushout complement of this boundary embedding, we
notice that there is only one solution to the re-pairing problem, and
that this solution is not plane. In a setting where all embeddings are
plane this is an unwanted case.

However in this case, solving the re-pairing problem and computing the
pushout complement already alerts us to the problem, since this graph
is not plane.  Notice first that the graph $G$ not carrying a rotation
system itself.  The circle is seemingly plane, but with the flags in $L$
fixed, there is no way this circle can be drawn on the plane without
edges crossing.
\begin{center}
\includegraphics[width=0.15\textwidth]{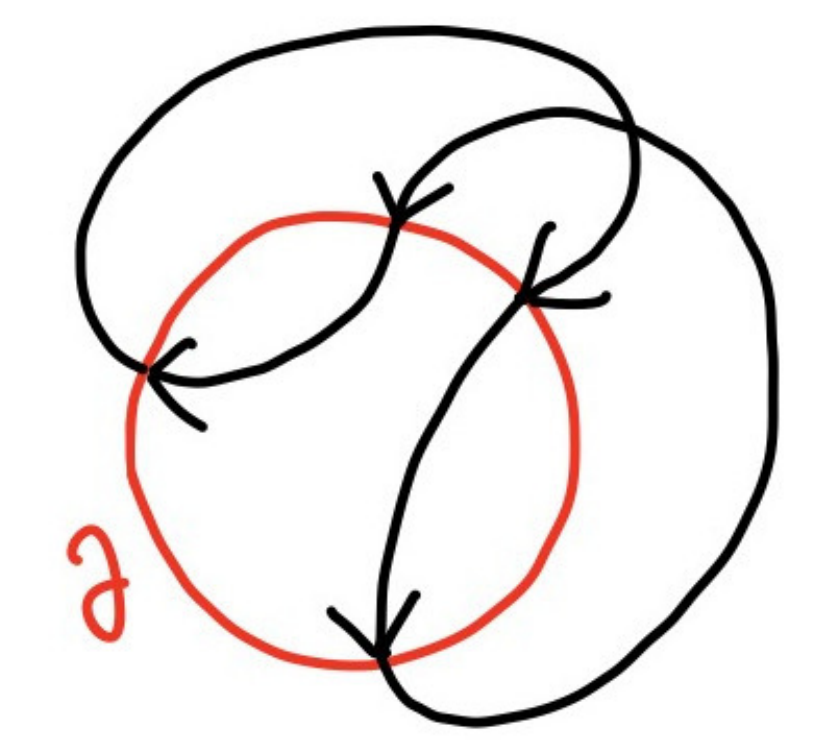}
\end{center}
Secondly, observe that the right hand side of the rewrite plays no
role: all the toplogical information is in the boundary
embedding. Thus we have a checkable condition to detect when a rewrite
will fail to preserve the surface.  We might hope for a necessary and
sufficient condition, or a stronger result allowing us to compute how
matching a given boundary into a graph alters the surface it is
embedded in.





\section{Conclusions and Further Work}
\label{sec:further-work}

In this work we have made some significant progress towards a purely
combinatorial formalisation of surface embedded strings diagrams.
Along the way we have introduced a new representation for symmetric
string diagrams and PROPs which removes several annoyances of
earlier approaches.

An obvious next step, already underway, is to formalise string
diagrams using the graph representation described here.  Unlike the
situation we have discussed in this paper, a morphism in a monoidal
category is not a closed surface -- it has a boundary, and it has
wires which impinge on that boundary.  Fortunately, the technology of
boundary vertices developed in Section~\ref{sec:dpo-rewr-suit} can be easily
adapted for this purpose.  At this point one could generalise to the
situation of a diagram on a surface with multiple boundaries.

However to build a complete theory of diagrams on surfaces we must
address two major topological questions.  The first was already
described in Section \ref{sec:planarity}: the preservation of
planarity by rewrites.  The second was briefly mentioned in the
introduction: disconnected graphs.  Minimally we must record the
relationships between components and faces of other components, and
consider how these relationships change under rewrites.  Many other
details arise, such as the orientation of circles.

A much simpler modification to the theory would be to consider the
undirected case.  This is relatively easy, since undirected graphs can
be obtained by a forgetful functor from the directed ones.  However
some details also change.  For example the repairing problem has more
solutions in the undirected setting than the directed.  However we
expect no major difficulties here.

Finally, a computerised implementation of this representation would be
most helpful for experiments and applications.

\paragraph*{Acknowledgements}
We would like to thank Tim Ophelders for helpful remarks on the different
solutions of the re-pairing problem, and the anonymous reviewers for their
comments.


\small
\bibliography{all}

\normalsize


\appendix

\section{Proofs}
\label{app:proofs}


\subsection*{From Section \ref{sec:suit-categ-graphs}}

\paragraph{Lemma \ref{lem:flag-bijections-compose}}
Let $f:G\to H$ and $g:H\to J$ be flag bijections; then $g\circ f$ is a
flag bijection.
\begin{proof}
For flag injectivity, we assume injectivity of the flag maps induced by $f$ and
$g$. If $f_V$ is undefined, so is the flag map. Consider flags $(e,v)$ and
$(e',v')$ where $(f_E \times f_V)$ is defined, $v=s(e)$, $v'=s(e')$, and assume
$g_F(f_F(e,v)) = g_F(f_F(e',v'))$. Because $f$ is a flag surjection and defined
on the given flags, Equation~\ref{eq:flag-surjectivity} holds strictly on $v$ and
$v'$. Therefore we get: $f_E(e) = s_H(f_V(v))$ and $f_E(e') =
s_H(f_V(v'))$. This lets us apply flag injectivity of $g$ to get $f_F(e,v) =
f_F(e',v')$, and flag injectivity of $f$ to reach $(e,v) = (e',v')$. The
argument applies equally to the target map.

For flag surjectivity we assume lax commutation of
Equation~\ref{eq:flag-surjectivity} for $f$ and $g$ and show that the composite
diagram also commutes laxly. 
\begin{center}
\begin{tikzcd}
  V_G \arrow[d, "s_G^{-1}"'] \arrow[r, "f_V"] \arrow[rd, phantom, "\geq"
    description] 
& V_H \arrow[r, "g_V"] \arrow[d, "s_H^{-1}"'] \arrow[rd, phantom, "\geq"
    description] 
& V_J \arrow[d, "s_J^{-1}"] \\
  P(E_G) \arrow[r, "P(f_E)"'] 
& P(E_H) \arrow[r,"P(g_E)"'] 
& P(E_J)                   
\end{tikzcd}
\begin{tikzcd}
  V_G \arrow[d, "s_G^{-1}"'] \arrow[r, "f_V"] \arrow[rrd, phantom, "\geq"
    description] 
& V_H \arrow[r, "g_V"]
& V_J \arrow[d, "s_J^{-1}"] \\
  P(E_G) \arrow[r, "P(f_E)"'] 
& P(E_H) \arrow[r, "P(g_E)"'] 
& P(E_J)                   
\end{tikzcd}
\end{center}
In the case of either $f_V$ or $g_V$ being
undefined, the composite $(g_V \circ f_V)$ is also undefined and the diagram
commutes laxly immediately. If both $f_V$ and $g_V$ are defined, both
their diagrams commute strictly, and by diagram gluing, their composite does as well.
\end{proof}


\paragraph{Lemma \ref{lem:morphisms-compose}}
  Defining composition point-wise, the composite of two morphisms of
  graphs with circles is again such a morphism. Additionally, if both morphisms
  are embeddings, their composition is an embedding as well.
  \begin{proof}
    Let $f:G\to G'$ and $g:G'\to G''$ be two morphisms; then $g\circ f
    = ((g_{V'} \circ f_V), (g_{A'}\circ f_A))$; since composition of
    partial functions is associative, we need only check that the four
    properties of Definition \ref{def:graph-with-circles} are
    preserved.

    Conditions \ref{item:a-total} and \ref{item:v-inj} follow from the
    properties of partial functions, and condition \ref{item:flag-bijective}
    (which includes condition \ref{item:flag-surjective}) follows from
    Lemma~\ref{lem:flag-bijections-compose}.  Observe that
    \begin{align*}
      (g\circ f)_{O} & = [g_{EO},g_O]\circ (f_{OE} + f_O)\\
                      & =  (g_{EO}\circ f_{OE}) + (g_O\circ f_{O})\\
                      & =  (g_{EO}\circ \emptyset) + (g_O\circ f_O)\\
                      & = g_O \circ f_O
    \end{align*}
    hence $(g\circ f)_O$ is injective since $f_O$ and $g_O$ are, satisfying
    condition~\ref{item:o-injective}. Finally, by a similar argument we have
    \begin{align*}
      (g\circ f)_{EO} & = [g_{EO},g_O]\circ (f_E + f_{EO})\\
                      & =  (g_{EO}\circ f_E) + (g_O\circ f_{EO})\\
                      & =  (\emptyset \circ f_E) + (g_O\circ \emptyset)\\
                      & = \emptyset
    \end{align*}
    hence the remaining condition~\ref{item:oe-empty} is satisfied.
  \end{proof}


  \subsection*{From Section \ref{sec:dpo-rewr-suit}}

\paragraph{Theorem \ref{thm:pushouts}}
In \catgraf, pushouts of partitioning spans exist. Further, the maps into the
pushout are embeddings.

\begin{proof}
  The proof will proceed via several intermediate results.  First we
  will explicitly define the pushout candidate
  \begin{tikzcd}[cramped,sep=small]
    L \arrow[r, "m"'] & G  & \arrow[l, "g"] C
  \end{tikzcd} (Definition~\ref{def:pushout-candidate}), show the
  constructed object $G$ is a valid graph, (Lemmas
  \ref{lem:source-map-well-defd} and \ref{lem:edges-and-circles}),
  show that $m$ and $g$ are indeed embeddings in \catgraf
  (Lemma~\ref{thm:po-morphisms-are-embeddings}), and finally show that
  the required universal property holds in in \catgraf
  (Lemma~\ref{thm:po-universal-property}).  This suffices to prove the
  theorem.
\end{proof}

\begin{definition}\label{def:pushout-candidate}
  Given the partitioning span
  \begin{tikzcd}[cramped,sep=small]
    L & B  \arrow[l, "l"'] \arrow[r, "c"] & C
  \end{tikzcd}, 
  we define the \emph{pushout candidate}
  \begin{tikzcd}[cramped,sep=small]
    L \arrow[r, "m"'] & G  & \arrow[l, "c"] C    
  \end{tikzcd} as follows.
  
  We construct the underlying sets and functions by pushout in \Pfn,
  \begin{equation}\label{eq:pushouts-for-pushout}
    \begin{tikzcd}
      V_L \arrow[d, "m_V"'] \arrow[r,leftarrow, "l_V"]
      & \{\partial,\bar\partial\} \arrow[d, "c_V"]
      \\
      V_G \arrow[r,leftarrow, "g_V"'] \lpushout
      & V_C
    \end{tikzcd}    
    \qquad\qquad
    \begin{tikzcd}
      A_L \arrow[d, "m_A"'] \arrow[r,leftarrow, "l_A"]
      & E_B \arrow[d, "c_A"]
      \\
      A_G \arrow[r,leftarrow, "g_A"'] \lpushout & A_C
    \end{tikzcd}
  \end{equation}
  so explicitly we have
  \[
    V_G = (V_C + V_L) \setminus V_B
    \qquad
    A_G = (A_L + A_C)/{\sim}
  \]
  where $\sim$ is the least equivalence relation such that
  $l_E(e) = c_E(e)$ for $e \in E_B$.  Next we define the source map by
  \begin{equation}\label{eq:pushout-src-map}
    s_G(e') = \left\{
      \begin{array}{l}
        s_L(e),  \text{ if } e' = m_A(e)
                  \text{ and } s_L(e)
                  \text{ is defined and } s_L(e) \neq \partial\\        
        s_C(e),  \text{ if } e' = g_A(e)
                  \text{ and } s_C(e)
                  \text{ is defined and } s_C(e) \neq \bar\partial\\
        \text{undefined otherwise.}
      \end{array}
    \right.
  \end{equation}
  for all $e' \in A_G$.  The target map $t_G$ is defined similarly.
  (Strictly speaking we have defined $s$ and $t$ on all of $A$; they
  will be restricted to $E$ when we have defined that.) 
  Finally we divide the arcs into edges and circles by setting
  \begin{align}\label{eq:pushout-edges-v-circles}
    E_G & = \{ e \in A_G : \text{ both } s_G(e) \text{ and } t_G(e)
    \text{ are defined }\}
    \\
    O_G & = A_G \setminus E_G
  \end{align}
\end{definition}

There are two properties that need to be checked to ensure that
the definition above yields a valid graph.  The source and target maps
should be well-defined partial functions; and all arcs should either
have two end points (\ie they are edges) or none (they are circles).

\begin{lemma}\label{lem:source-map-well-defd}
  Equation~\eqref{eq:pushout-src-map} defines a partial function: if
  $s_G(e')$ is defined, it is single-valued.
  \begin{proof}
    There are two things to check.  First we show that if the first or
    second clause of the definition applies it is single valued.  We
    then show that at most one of those clauses can apply.

    Suppose that in $L$ we have distinct $e_1$, $e_2$ such that
    $m_A(e_1) = m_A(e_2)$ and $s_L(e_1)\neq \partial$.  Since they are
    distinct in $L$ and identified in $G$, we must have distinct
    $e_1,e_2 \in B$ such that $c_A(e_1) = c_A(e_2)$ in $C$.  By
    Lemma~\ref{lem:self-loop-creation} this gives a self-loop at
    $\bar\partial$ in $C$, which in turn implies that
    $s_L(e_2) = \partial$.  Hence $L$ provides at most one candidate
    source vertex for every edge in $G$, and a similar argument can be
    made for $C$.

    Now suppose $m_A(e_1) = g_A(e_2)$, and that
    $s_L(e_1) \neq \partial$ and $s_C(e_2) \neq \bar\partial$.  Since
    the edges are identified in $G$ they are both present in $B$.
    Since $s_L(e_1) \neq \partial$ we have $s_B(e_1) = \bar\partial$,
    from which $s_C(e_1) = \bar\partial$.  Since
    $s_C(e_2) \neq \bar\partial$, $e_1$ and $e_2$ are distinct in $C$.
    Therefore we must have $e_1$ and $e_2$ identified in $L$;
    therefore, by Lemma~\ref{lem:self-loop-creation}, $e_1$ must be a
    self-loop at $\partial$ which contradicts our original assumption.
    Therefore there is at most one candidate source vertex and the map
    $s_G$ is well defined in (\ref{eq:pushout-src-map}).  
  \end{proof}
\end{lemma}

\noindent
The preceding argument applies equally to the target map $t_G$.

\begin{lemma}\label{lem:path-in-B}
  Let \catP be the pairing graph of the partitioning span
  \begin{tikzcd}[cramped,sep=small]
    L & B  \arrow[l, "l"'] \arrow[r, "c"] & C
  \end{tikzcd}, and let $G$ be its pushout candidate.
  \begin{enumerate}
  \item Suppose $e$ and $e'$ are edges in $B$; if $e$ and
    $e'$ are in the same component of \catP then
    $(m_A \circ l_E)(e) = (m_A \circ l_E)(e')$.
  \item Let $e$ be any arc in $A_G$; then its preimage in $B$ is
    either empty or is exactly one connected component of \catP.
  \end{enumerate}
  \begin{proof}
    (1)  Suppose that $e$ and $e'$ are the same component of \catP.
    We use induction on the length of the path from $e$ to $e'$ in
    \catP.  If the path is length zero, then $e = e'$ and the property
    holds trivially.  Otherwise, let $e''$ be the predecessor of
    $e'$.  By induction, and (\ref{eq:pushouts-for-pushout}), we have
    \[
      (m_A \circ l_E)(e'')
      = (m_A \circ l_E)(e)
      = (g_A \circ c_E)(e)
      = (g_A \circ c_E)(e'')
    \]
    Since $e'$ and $e''$ are adjacent in \catP we must have either
    $l_E(e')= l_E(e'')$ or  $c_E(e')= c_E(e'')$ depending on the
    colour of the edge.  From this the result follows.

    (2) Let $e\in A_G$ and suppose that $e_1 \in (m_A\circ
    l_E)^{-1}(e)$ in $B$.  Either $e_1$ is a component on its own, or
    it has a neighbour $e_2$.  By the definition of \catP either
    $l_E(e_1) = l_E(e_2)$ or $c_E(e_1) = c_E(e_2)$ depending on the
    colour of the edge.  Therefore we have
    \[
      (m_A\circ l_E)(e_2)
      = (m_A\circ l_E)(e_1)
      = e
    \]
    so $e_2$ is also in the pre-image of $e$.  By induction, the
    entire component containing $e_1$ must also be included in the
    pre-image. 

    For the converse, recall that $A_G = (A_L + A_C)/{\sim}$ where
    $\sim$ is the least equivalence relation such that
    $l_E(e_i) = c_E(e_i)$ for $e_i \in E_B$.  Therefore if distinct
    $e'$ and $e'' \in E_B$ both belong to the preimage of $e \in A_G$,
    there necessarily exists a chain of equalities
    \[
      l_E(e') = l_E(e_1),\quad
      c_E(e_1) = c_E(e_2),\quad
      l_E(e_2) = l_E(e_3),\quad      
      \ldots ,\quad
      c_E(e_n) = c_E(e'')
    \]
    to place them in the same equivalence class.  Such a chain of
    equalities precisely defines a path from $e'$ to $e''$ in \catP,
    hence if two edges of $B$ are identified in the pushout, they
    belong to the same component in the pairing graph.
  \end{proof}
\end{lemma}

\begin{lemma}\label{lem:edges-and-circles}
  Let $G$ be the pushout candidate defined above.  For all arcs $e \in
  A_G$ either both $s_G(e)$ and $t_G(e)$ are defined or neither is.
  \begin{proof}
    Consider the preimage of $e$ in $B$; if it is empty then
    $e$ is simply included in $G$ from either $L$ or $C$, along with
    both its end points.

    Otherwise, by Lemma \ref{lem:path-in-B}, $e$ corresponds to a
    connected component $p$ of the pairing graph \catP.  By corollary
    \ref{cor:pgraph-comps-are-paths-in-B} such components can be
    either line graphs or closed loops.  If $p$ is a closed loop, for
    all $e_i \in p$ we have
    \[
      s_L(l_E(e_i)) = t_L(l_E(e_i)) = \partial
      \qquad\text{ and }\qquad
      s_C(c_E(e_i)) = t_C(c_E(e_i)) = \bar\partial
    \]
    so, by (\ref{eq:pushout-src-map}), neither $s_G(e)$ nor $t_G(e)$
    is defined.  If, on the other hand, $p$ forms a path $e_1, e_2,
    \ldots, e_n$, its ends provide the source and target.
    Specifically, if $e_1$ positive in \catP then $s_C(c_E(e_1)) \neq
    \bar\partial$ and if it is negative $s_L(l_E(e_1)) \neq
    \partial$; if $e_n$ is positive $t_L(l_E(e_n)) \neq
    \partial$, and if $e_n$ is negative $t_C(c_E(e_n)) \neq
    \bar\partial$.

    Hence $s_G(e)$ is defined if and only if $t_G(e)$ is defined.
    Therefore the division of $A_G$ into edges and circles is correct
    and $G$ is indeed a valid graph.
%
 %
  %
  \end{proof}
\end{lemma}


\begin{lemma}\label{thm:po-morphisms-are-embeddings}
  The arrows of the cospan   \begin{tikzcd}[cramped,sep=small]
    L \arrow[r, "m"'] & G  & \arrow[l, "g"] C
  \end{tikzcd} defined by the pushout candidate are embeddings in
  \catgraf. 
  \begin{proof}
    We will show the result for $m$; the proof for $g$ is the same.
    Note that Properties \ref{item:v-inj} and \ref{item:a-total} are
    automatic from the underlying pushouts in \Pfn. Since the graph
    $B$ has no circles, the $m_O$ component is injective by
    construction (Property \ref{item:o-injective}) and since no arc
    gets a source or target in $G$ unless its preimage had one, the
    component $m_{OE}$ is empty as required (Property
    \ref{item:oe-empty}).  Finally we have to show that the induced
    map $(m_V,m_E)$ is a flag bijection.  First note that if $m_E(e)$
    is undefined then $e$ is necessarily a self-loop at $\partial$,
    and $m_V(\partial)$ is always undefined, so the squares
    (\ref{eq:edge-lax}) commute.  Otherwise if $(f_V \circ s_L)(e)$ is
    defined then the square commutes directly by the definition of
    $s_G$ above, and similarly for $t_G$.  Finally for all
    $v \neq \partial \in V_L$, we have that $m_V(v)$ is defined. By
    the definition of $s_G$ and $t_G$, $e$ is a flag at $v$ if and
    only if $m_E(e)$ is a flag at $m_V(v)$.  Flag injectivity and flag
    surjectivity follow immediately.  Hence $m$ is an embedding in
    \catgraf.
  \end{proof}
\end{lemma}
    
\begin{lemma}\label{thm:po-universal-property}
  the cospan 
  \begin{tikzcd}[cramped,sep=small]
    L \arrow[r, "m"] & G & \arrow[l, "g"'] C
  \end{tikzcd}
  has the required universal property.
  \[
    \begin{tikzcd}
      & L \arrow[d, "m"']
          \arrow[r,leftarrow, "l"]
          \arrow[ddl,"m'"',bend right]
      & B \arrow[d, "c"]
      \\
      & G \arrow[r,leftarrow, "g"']
          \arrow[dl,dashed,"f"]
          \lpushout            
      & C \arrow[dll,"g'",bend left]
      \\
      G'
    \end{tikzcd}    
  \]
  \begin{proof}
    Since the underlying sets and functions are constructed via
    pushout the required mediating map $f = (f_V,f_A)$ exists; we need
    to show that it is a morphism of \catgraf.  Property
    \ref{item:a-total} follows from $m'$ and $g'$ satisfying it as
    well. For the $f_{OE}$ to be empty (Property \ref{item:oe-empty}),
    use the fact that $m'_{OE}$ and $g'_{OE}$ are empty for circles in
    $L$ and $C$, because they are morphisms in \catgraf. The remaining
    case for a circle to appear in $G$ is as the pushout of some edges
    in $B$ being identified in one instance of the re-pairing
    problem. In this case, because the outer square has to commute for
    the edge component, these edges have to be identified, and hence
    form a circle, in $G'$, too. This makes $f_{OE}$ empty. For flag
    surjectivity between the underlying graphs (Property
    \ref{item:flag-surjective}), observe that the vertex set $V_G$ is
    the disjoint union of vertex sets $V_L$ and $V_C$. Because $m'$
    and $g'$ are valid morphisms in $\catgraf$, they are flag
    surjective, and therefore so is $f$.
  \end{proof}
\end{lemma}

This completes the proof of Theorem~\ref{thm:pushouts}.


\paragraph{Lemma \ref{lem:repairing-has-a-solution}}
Given a boundary embedding 
\begin{tikzcd}[cramped,sep=small]
B \arrow[r,"l"] & L \arrow[r,"m"] & G 
\end{tikzcd} 
a solution to the re-pairing problem always exists, but it is not
necessarily unique.
\begin{proof}
  Note that any half-pairing graph has connected components of at most
  two vertices, linked by a (blue) edge from a positive vertex to a
  negative one.  Define the component of an arc by
  \[
    k(a) =  (m_A \circ l_A)^{-1}(a) \text{ for all } a \in A_G
  \]
  Note that this defines a partition of the set
  $E_B \simeq \sum_{a\in A_G} k(a)$, and each (non-empty) $k(a)$
  determines a connected component of the solution to the re-pairing
  problem.  We'll abuse notation and use $k(a)$ to also denote the
  subgraph of the half-pairing graph whose vertices are $k(a)$.
  There are two cases depending whether $a$ is a circle or an
  edge.
  \begin{enumerate}
  \item Suppose $a\in O_G$; we form a closed loop involving all
    $e\in k(a)$, by adding red edges as follows.  Pick a degree-one
    positive vertex $p$ follow the incident blue edge to the negative
    vertex $n$; now pick another a degree-one positive vertex $p'$
    which is not connected to $n$.  Add a red edge from $n$ to $p'$.
    Repeat the process starting from $p'$.  When no more vertices
    remain, close the loop by adding a red edge from the final
    negative vertex back to $p$.  Since $a$ is a circle, $k(a)$
    necessarily contains an even number of vertices, so closing the
    loop is always possible.
  \item The case when $a$ is an edge is slightly more complex because
    edges have end points; $k(a)$ may contain zero, one, or two
    degree-zero vertices depending how many of its end points are
    defined by vertices in $L$.  We will connect the vertices as
    previously, but in a line, rather than a loop.  Since we can only
    add red edges, and only one at each vertex, the degree-zero
    vertices will necessarily be the end points of this line.
  \end{enumerate}
\end{proof}

\newpage
\section{Examples}
\label{app:examples}

This collection captures some of the corner cases we do or do not want to allow
as morphisms in the category of graphs with circles as described in
Definition~\ref{def:graph-with-circles}.
\begin{figure}[!h]
\centering
\includegraphics[width=.45\textwidth]{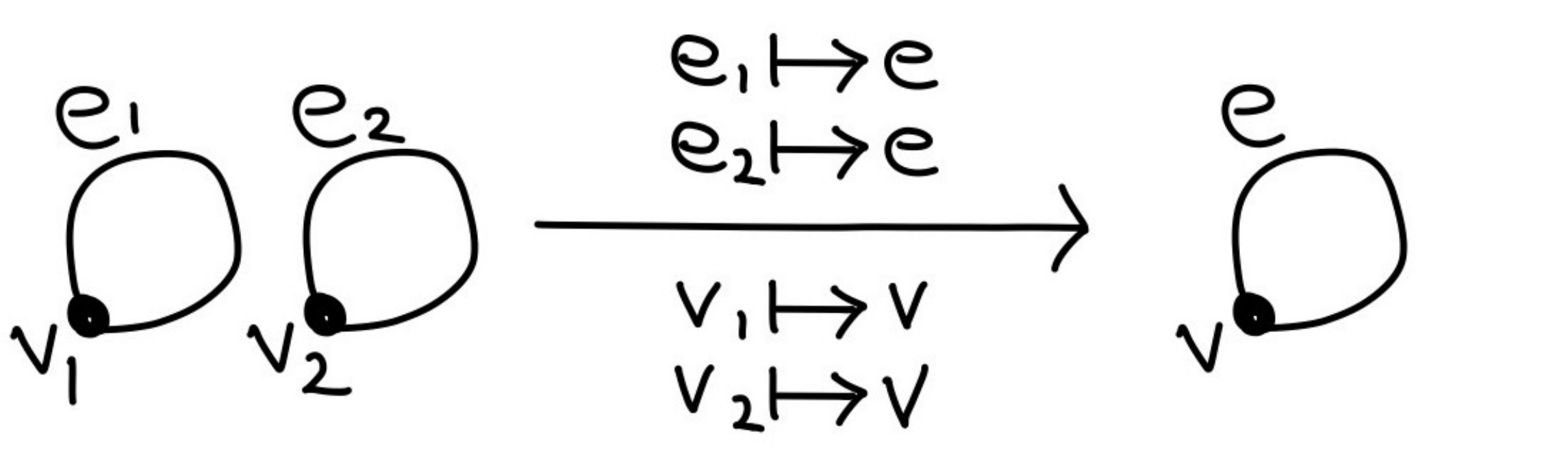}
\caption{A flag surjective, but not flag bijective map. This is a valid
of \catgraf.}
\label{fig:inj1}
\end{figure}

\begin{figure}[!h]
\centering
\includegraphics[width=.45\textwidth]{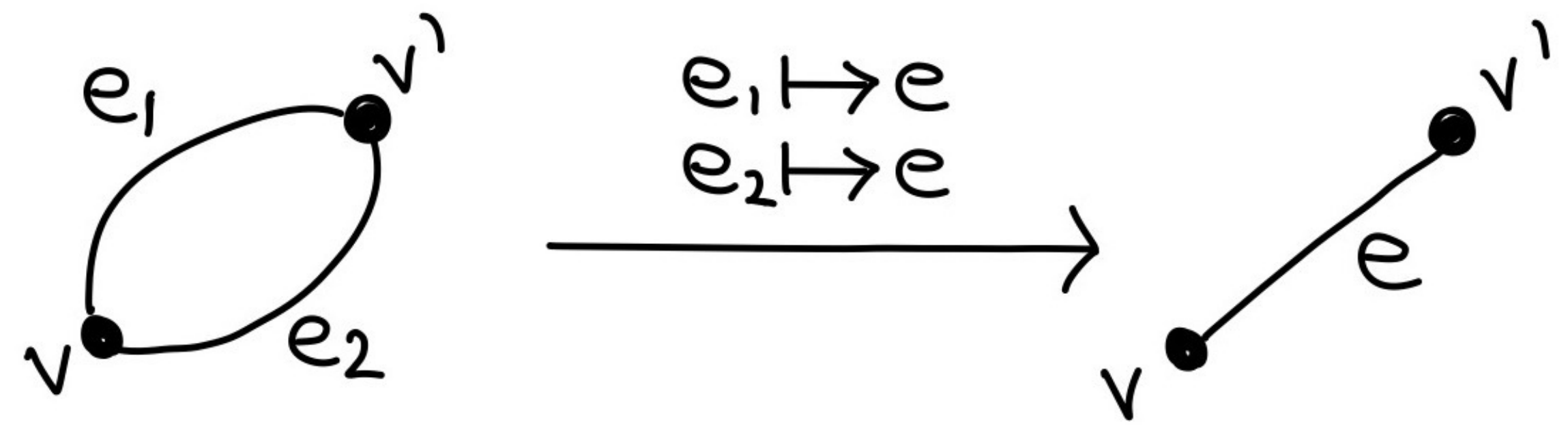}
\caption{An example of a flag surjective but not flag bijective morphism of \catgraf.}
\label{fig:inj2}
\end{figure}

\begin{figure}[!h]
\centering
\includegraphics[width=.45\textwidth]{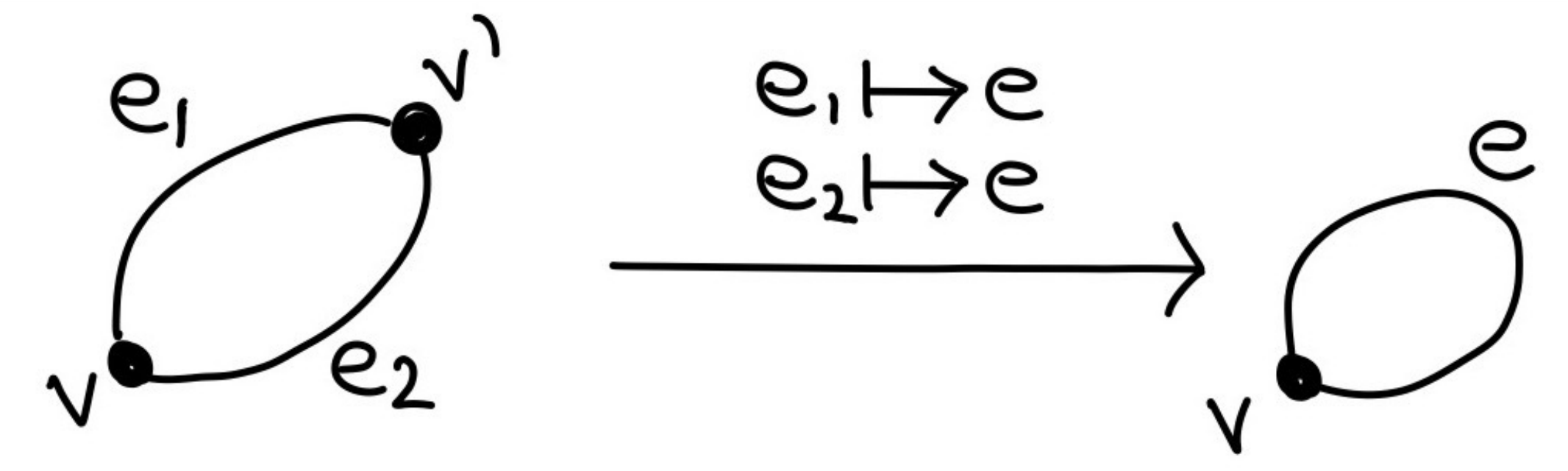}
\caption{An example of an embedding (hence a flag bijective morphism) in \catgraf.}
\label{fig:inj2b}
\end{figure}

\begin{figure}[!h]
\centering
\includegraphics[width=.35\textwidth]{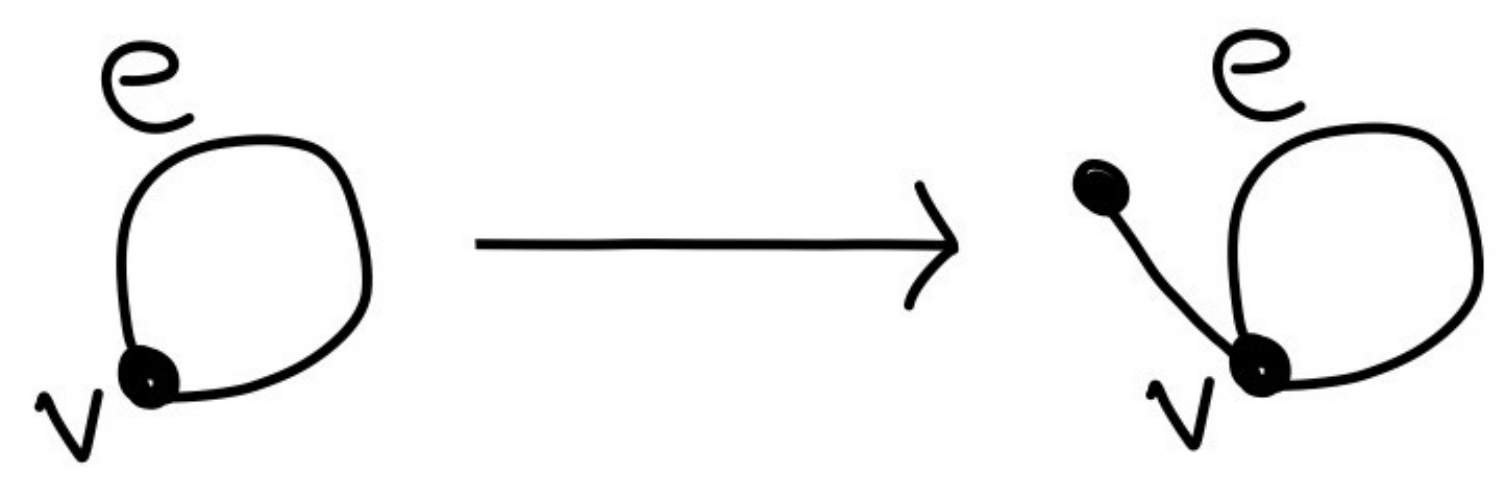}
\caption{This map is not flag surjective and therefore not a valid morphism
in \catgraf.}
\label{fig:surj}
\end{figure}

\begin{figure}[!h]
\centering
\includegraphics[width=.3\textwidth]{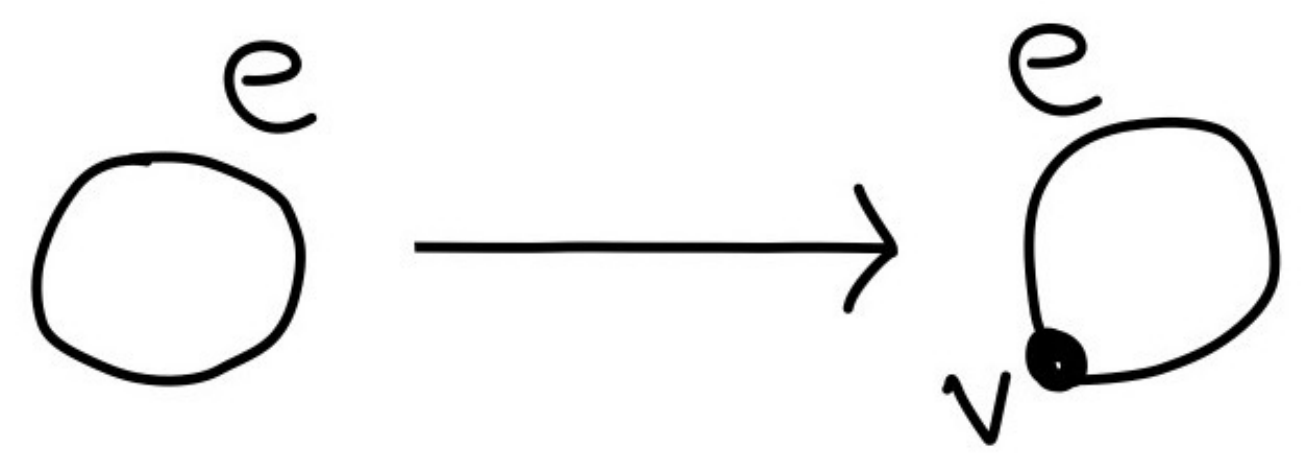}
\caption{This example is not a valid morphism in \catgraf because it does not
respect Condition~\ref{item:oe-empty}.}
\label{fig:circle}
\end{figure}

\end{document}